\documentclass[aps,pre,preprint,superscriptaddress,showpacs,showkeys,preprintnumbers]{revtex4-1}

\newtheorem{proposition}{Proposition}
\newtheorem{lemma}{Lemma}

\newtheorem{remark}{Remark}

\usepackage{graphicx}
\usepackage{bm}        
\usepackage{amssymb}   

\begin{document}


\title{Collapse of generalized Euler and surface quasi-geostrophic point-vortices}



\author{Gualtiero Badin}
\affiliation{Center for Earth System Research and Sustainability (CEN), University of Hamburg, Hamburg, Germany}
\email[]{gualtiero.badin@uni-hamburg.de}

\author{Anna M. Barry}
\affiliation{Department of Mathematics, The University of Auckland, New Zealand}
\email[]{anna.barry@auckland.ac.nz}


\date{\today}

\begin{abstract}
Point vortex models are presented for the generalized Euler equations, which are characterized by a fractional Laplacian relation between the active scalar and the streamfunction.  Special focus is given to the case of the surface quasi-geostrophic (SQG) equations, for which the existence of finite-time singularities is still a matter of debate. Point vortex trajectories are expressed using Nambu dynamics.  The formulation is based on a noncanonical bracket and allows for a geometrical interpretation of trajectories as intersections of level sets of the Hamiltonian and Casimir. Within this setting, we focus on the collapse of solutions for the three point vortex model. In particular, we show that for SQG the collapse can be either self-similar or non-self-similar. Self-similarity occurs only when the Hamiltonian is zero, while non-self-similarity appears for non-zero values of the same. For both cases, collapse is allowed for any choice of circulations within a permitted interval. These results differ strikingly from the classical point vortex model, where collapse is self-similar for any value of the Hamiltonian, but the vortex circulations must satisfy a strict relationship. Results may also shed a light on the formation of singularities in the SQG partial differential equations, where the singularity is thought to be reached only in a self-similar way.
\end{abstract}


\maketitle


\section{Introduction}



In this study we are concerned with an incompressible fluid in 
${\textbf R}^2$, with motion governed by the equation, here written in local coordinates,
\begin{equation}
\frac{\partial \zeta}{\partial t} + \frac{\partial(\psi, \zeta)}{\partial (x,y)} = 0~,
\label{eq:alpha1}
\end{equation}
where $\zeta(x,y,t)$ is an active scalar and $\psi(x,y,t)$ is the streamfunction of the flow, which satisfy the relationship
\begin{equation}
- \zeta = (- \Delta)^{\alpha /2}\psi~,
\label{eq:alpha2}
\end{equation}
with parameter $\alpha \in {\textbf R}$.

The streamfunction is related to the horizontal velocity $(u,v)$ via
\begin{equation}
(u,v) = \left( - \frac{\partial \psi}{\partial y} , \frac{\partial \psi}{\partial x} \right) ~.
\label{eq:alpha3}
\end{equation}
We define $\nabla^{\perp} = (- \partial / \partial y  ,~ \partial / \partial x )$ and $\Lambda = (- \Delta )^{\alpha /2}$, so that the velocity can be expressed in terms of the active scalar:
\begin{equation}
(u,v) = \nabla^{\perp} \Lambda^{-1} \zeta = ( - \mathcal{R}_2 \zeta , \mathcal{R}_1 \zeta)~,
\label{eq:alpha4}
\end{equation}
where $\mathcal{R}_1,~\mathcal{R}_2$ denote the Riesz transforms. These equations take the name of \emph{generalized Euler equations} or \emph{$\alpha$-models} (note that one should not confuse the $\alpha$-models with the Euler-$\alpha$ equations, that arise instead as a Lagrangian average of the fluctuations acting on the Euler equations, \cite{holm1998euler}).

In these models,  $\alpha = 2$ corresponds to the widely studied Euler equations with the vorticity taking the place of the active scalar, while $\alpha < 2$ and $\alpha > 2$ correspond to so-called \textit{local} and \textit{nonlocal} dynamics respectively.  In particular, note that increasing $\alpha$ serves to weaken the local coupling between the active scalar and the streamfunction.


The turbulence emerging in these models has been studied for a variety of values of $\alpha$ (see e.g.  \cite{pierrehumbertetal94,smith2002turbulent,tran2002constraints,tran2004nonlinear,tran2010effective,schorghofer2000universality,burgess2013spectral,burgess2015kraichnan,venaille2015violent,foussard2017relative,conti18arxiv}).  In the present article, we focus on the surface quasi-geostrophic (SQG) model \cite{blumen78,heldetal95,lapeyre17,badin2018variational} obtained when $\alpha = 1$. 
This model emerges in the study of atmospheric and oceanic dynamics when the active scalar is given by the temperature at one of the boundaries (e.g. the atmospheric tropopause or the oceanic surface), with the potential vorticity in the interior of the fluid set to zero.
In this case, the emerging turbulence is characterized by a forward (i.e., toward large wavenumbers) cascade of kinetic energy \cite{tulloch2006theory,capet2008surface}, which might make it a candidate for the route to dissipation of geophysical flows, as well as by a forward cascade of temperature variance.  The latter results in the formation of fronts, which in turn are important for the mixing of passive tracers \cite{scott2006local}. In the ocean, SQG dynamics may shed light on the formation of submesoscale dynamics  \cite{McWilliams2016}, which again are important for the mixing of passive tracers \cite{badin2011lateral,Shcherbina2015,mukiibi2016three}. For a study on the relationship between quasi geostrophic (QG) and SQG turbulence, see e.g. \cite{badin2014role}. The stability of SQG vortices was investigated in e.g. \cite{carton2009instability,dritschel2011exact,harvey2011perturbed,harvey2011instability,bembenek2015realizing,carton2016vortex,Badinpoulin2018}.
 
In mathematical terms, the SQG model, which is 2D, shows strong analogies with the 3D Euler equation \cite{constantinetal94}, for which the existence of finite-time singularities is still a matter of debate. These observations sparked much interest because they suggest that the study of the regularity of the SQG model could provide hints for the formation of singularities in the 3D Euler equation, see e.g. \cite{constantinetal94,constantin1994singular,majda1996two,ohkitani1997inviscid,constantin1998nonsingular,constantin1999behavior,cordoba2002growth,cordoba2002scalars,cordoba2004maximum,rodrigo2004vortex,cordoba2005evidence,rodrigo2005evolution,wu2005solutions,deng2006level,dong2008finite,ju2006geometric,li2009existence,marchand2008existence,marchand2008weak,scott2011scenario,constantin2012new,scott2014numerical}. Amongst these studies, the numerical work reported in \cite{ohkitani2012asymptotics} shows that the $\alpha$-models (\ref{eq:alpha1})-(\ref{eq:alpha4}) possess a value of $\alpha$ in the interval $[0,2]$ for which the solutions behave in the most singular manner.  Further, both heuristic and mathematical arguments suggest that if the SQG equations are singular, the formation of singularities must be self-similar, see e.g. \cite{hoyer1982closure,pierrehumbertetal94,heldetal95,scott2011scenario,scott2014numerical}.

It is well known since Helmholtz \cite{helmholtz1858} that point vortex models can be derived from the Euler equations. By dividing the fluid into a number of separated regions with arbitrarily small area and considering the limit, each vortex is seen to approach a single point with infinite vorticity and finite circulation (e.g. \cite{aref2007point,marchioro2012mathematical,newton2013n}). Of particular interest is the case given by three vortices, which possesses many non-trivial properties but is completely integrable \cite{synge1949motion}. One particularity of point-vortex dynamics is that, if the system of point vortices satisfy some necessary constraints on their circulations, there exist motions where the vortices \emph{collapse} to a point in a finite time \cite{aref1979motion,novikov1979vortex}. This collapse is subject to the initial conditions of the system and can take place for an arbitrary number of vortices \cite{novikov1979vortex,o1987stationary}. For 2D Euler dynamics, the collapse must be self-similar,  i.e. the motion takes place without change of shape of the system \cite{aref2010self}. On the sphere and for 4 degrees of freedom, \cite{sakajo2008non} showed that point vortices can undergo non-self-similar collapse. Further, the spherical geometry allows also for partial collapse, in which only a subset of the point vortices coalesce. For further studies on collapse of point vortices, see  \cite{tavantzis1988dynamics,kimura1990vortex,kimura1990parametric,kimura1991vortex,leoncini2000motion,leoncini2001chaotic,hernandez2007collisions,sakajo2012instantaneous,sakajo2013anomalous,kudela2014collapse}. 

The Hamiltonian nature of the point-vortex model allows for the use of powerful machinery built in classical mechanics \cite{kirchhoff1876vorlesungen}. A notable alternative to the Hamiltonian formulation is due to Nambu \cite{Nambu1973}. The latter is based on Liouville's Theorem and relies not only on the Hamiltonian but also on an arbitrary number of Casimirs, i.e. the singularities of the Poisson operator, of the system. The resulting dynamics is determined by a skew-symmetric bracket, known as the Nambu bracket.
For finite-dimensional systems (e.g. \cite{NevirBlender1994}), the resulting dynamics has an elegant geometric interpretation: dynamical motions follow intersections of the manifolds defined by level-sets of the Hamiltonian and of the Casimirs. 

Following \cite{blender2015hydrodynamic}, consider a system with three degrees of freedom ${\textbf r}=(r_1,r_2,r_3)$. 
If the system possesses two invariants, namely the Hamiltonian $H = H({\textbf r})$ and the Casimir $C = C({\textbf r})$, one has
\begin{eqnarray}
\dot{H}  & = & \nabla_{r} H \cdot \dot{{\textbf r}} = 0 ~, 
\label{eq:21} \\
\dot{C}  & = & \nabla_{r} C \cdot \dot{{\textbf r}} = 0~,
\label{eq:22}
\end{eqnarray}
where $\nabla_{r}$ indicates the gradient operator in ${\textbf r}$ space. 
The system (\ref{eq:21})-(\ref{eq:22}) implies an orthogonality relationship between $\dot{{\textbf r}}$,  $\nabla_r H$  and  $\nabla_r C$, so that
\begin{equation}
\dot{{\textbf r}} = \nabla_r C \times \nabla_r H ,
\label{eq:23}
\end{equation}	
which shows that the flow is along the intersection of the manifolds defined by the level-sets of $H$ and $C$. Notice that in (\ref{eq:23}) time might be properly rescaled to include proportionality terms. Equation (\ref{eq:23}) is the canonical Nambu form
\begin{eqnarray}
\dot{r}_i & = & \frac{\partial \left( r_i , C, H\right)}{\partial (r_i, r_j, r_k )} \nonumber \\
& = & \varepsilon_{ijk} \frac{\partial C}{\partial r_j}\frac{\partial H}{\partial r_k}    \nonumber \\
& = & \left\{ r_i , C, H \right\}_N \qquad i,j,k=1,2,3
\label{eq:25}
\end{eqnarray}
where $\varepsilon_{ijk}$ is the anti-symmetric Levi-Civita symbol and where the curly braces indicate the (here canonical) Nambu bracket. 
The dynamics of an arbitrary state space function $F({\textbf r})$ is thus given by 
\begin{equation}
\dot{F} = \left\{ F,C,H\right\}_N ~.
\label{eq:26}
\end{equation}
If (\ref{eq:26}) is generalized as $\{F_1,...,F_n\}$, the bracket is a multilinear map
\begin{equation}
\left\{ ~,~,~\right\}_N ~:~[C^{\infty}(X)]^{\otimes n} \rightarrow C^{\infty}(X),
\label{eq:27a}
\end{equation}
for all $F_i~(i=1,...,n) \in X$, where $X$ is a smooth manifold. Equation (\ref{eq:27a}) satisfies the following properties:
\begin{itemize}
\item {\it Skew-symmetry} 
\begin{equation}
\left\{ F_1,\cdots,F_n\right\}_N =(-1)^{\varepsilon (\sigma)} \left\{ F_{\sigma_1},\cdots,F_{\sigma_n}\right\}_N,
\label{eq:27b}
\end{equation}
where $\varepsilon  (\sigma)$ is the parity of a permutation $\sigma$.
\item {\it Leibniz Rule} 
\begin{equation}
\left\{ F_1,\cdots,F_n\right\}_N =F_1 \left\{ F_2,\cdots,F_n\right\}_N +\left\{ F_1,F_3,\cdots,F_n\right\}_N F_2.
\label{eq:27c}
\end{equation}
\item {\it Jacobi identity} 
\begin{eqnarray}
\{\{F_1,...,F_{n-1},F_n\}_N,F_{n+1},...,F_{2n-1}\}_N +\\ \nonumber
+\{F_n,\{F_1, ...,F_{n-1},F_{n+1}\}_N,F_{n+2},...,F_{2n-1}\}_N + \cdots \\ \nonumber
+ \{F_n,...,F_{2n-2},\{F_1,...,F_{n-1},F_{2n-1}\}_N \}_N  \\ \nonumber
= \{F_1,...,F_{n-1},\{F_n,...,F_{2n-1}\}_N \}_N~.
\label{eq:27d}
\end{eqnarray}
\end{itemize}
For more algebraic properties of Nambu brackets see \cite{Takhtajan1994}.

The extension of Nambu mechanics to infinite-dimensional systems was performed in \cite{NevirBlender1993}, where the Nambu brackets for fluid dynamics were formulated using enstrophy and helicity as Casimirs in 2D and 3D respectively. In geophysical fluid dynamics, the Nambu brackets have been used e.g. by  \cite{Salmon2005, bridges2006numerical,Salmon2007,Bihlo2008,GassmannHerzog2008,NevirSommer2009,SommerNevir2009,SalazarKurgansky2010,BlenderLucarini2013,blender2017viscous,blender2017construction}. Of particular interest for this study is the derivation of the Nambu brackets for the $\alpha$-models, derived for the first time in \cite{blender2015hydrodynamic}.  Given a 2D continuous system with the dynamic variable  $\zeta$ as previously defined, along with the Hamiltonian $H = H[\zeta]$ and Casimir $C = C[\zeta]$, one has the generalized Euler equation
\begin{equation}
\dot{ \zeta} + \frac{ \partial \left( C_{\zeta}, H_{\zeta} \right) }{\partial (x , y) } = 0~,
\label{eq:36}
\end{equation}
where the the subscripts indicate variational derivatives with respect to $\zeta$. Equation (\ref{eq:36}) is equivalent to (\ref{eq:alpha1}). Arbitrary functionals $F[\zeta]$  evolve according to
\begin{equation}
\dot{ F} + \{ F,H,C \}_N=0 ,
\label{eq:37}
\end{equation}
with the noncanonical Nambu bracket 
\begin{equation}
\{F,H,C\}_N = - \int_A F_{\zeta} \frac{ \partial \left( H_{ \zeta}, C_{ \zeta} \right)}{\partial (x , y)} dA .
\label{eq:38}
\end{equation}

The classical ($\alpha=2$) point-vortex model was first studied using Nambu mechanics in \cite{Makhaldiani07,Makhaldiani12}. In particular, the work \cite{MuellerNevir14} used geometric properties of finite-dimensional Nambu mechanics to analyse collapse of three point-vortices. This study looked for intersections of level-sets of the Hamiltonian and Casimirs which pass through the collapse point, i.e. where the mutual distances between the vortices is zero. The result represents a manifold of necessary conditions for collapse.

While the analyses listed so far focus on the classical 2D Euler model, the results can be generalized to other models. For example, SQG point vortices have been studied in \cite{taylor2016dynamics}, while interactions between SQG and QG point vortices are studied in \cite{lim2001point}. A number of questions thus arise: Does the SQG point-vortex model possess collapse solutions? If yes, must collapse be self-similar? These questions, beyond being interesting per se, may also be of interest due to the following motivation. It is believed that the formation of singularities in SQG should happen, if at all, in a self-similar way. Thus the way in which SQG point vortices undergo collapse might pose an intriguing analogy for the study of singular solutions of the SQG PDE along the lines proposed in \cite{novikov1979vortex} for the Euler equation.

In the next section we outline the derivation of point vortex equations for the $\alpha$-models and consider relevant properties of the models.
In particular, we make a Nambu formulation and in Section 3 we use this to set up collapse conditions for the models.  Section 4 contains a detailed study of the SQG point vortex model ($\alpha=1$), establishing the existence of collapse solutions and analysing their self-similar structure. We conclude with a discussion in Section 5.
%
\section{Point vortex equations for the $\alpha$-models}
%
In this section we derive point vortex equations for the $\alpha$-models. When $\alpha > 3$, we will see that the effect of one vortex on another increases with distance, and hence this case is considered unphysical. Because of this,
only the interval $\alpha \, \in \, ]0,3]$ will be considered. Note that this interval includes the transition between local ($\alpha < 2$) and nonlocal ($\alpha > 2$) dynamics. The particular case of SQG will first be considered in Section \ref{sec:sqg}.

\subsection{Green's functions}

Consider a point vortex with circulation $\Gamma$ placed at the origin, so that
\begin{equation}
\zeta ({\textbf r} ) = \Gamma \delta ({\textbf r})~.
\label{eq:GF1}
\end{equation}  
The Green's function for the $\alpha$-models is
\begin{equation}
G^{(\alpha)} ({\textbf r} ) = - \left( - \Delta \right)^{- \alpha / 2} \delta ({\textbf r})~,
\label{eq:GF2}
\end{equation}
from which the streamfunction can be calculated as
\begin{equation}
\psi ({\textbf r} ) = \int G^{(\alpha)} ({\textbf r}, {\textbf r'} ) \zeta ( {\textbf r'} ) d {\textbf r'}~.
\label{eq:GF3}
\end{equation}
The form for (\ref{eq:GF2}) can be found by taking its Fourier transform \cite{Iwayama2010}. Calculations give
\begin{equation}
G^{(\alpha)} ({\textbf r} )  =  \Psi (\alpha) {\textbf r}^{\alpha - 2}~, \qquad \alpha  \neq 2~, 
\label{eq:GF4b}
\end{equation} 
where 
\begin{equation}
\Psi (\alpha ) = - \left\{ 2^{\alpha} \left[ \Gamma \left( \frac{ \alpha }{2} \right) \right]^2 \sin \left( \frac{\alpha \pi}{2} \right) \right\}^{-1}~.
\label{eq:GF5}
\end{equation}
As shown by \cite{Iwayama2010}, the function $\Psi (\alpha)$ has a discontinuity at $\alpha=2$. In this case, one has
\begin{equation}
G^{(2)} ({\textbf r} )  =  \frac{1}{2 \pi} \ln {\textbf r} + C~, 
\label{eq:GF4a} 
\end{equation} 
where $C$ is an arbitrary constant


\subsection{Equations of motion}

Given $n$ point vortices with circulations $\Gamma_i$, $i=1,...,n$, let 
\begin{equation}
r^2_{ij}= \left( x_i - x_j \right)^2 + \left( y_i - y_j \right)^2 ~,
\label{eq:r2}
\end{equation}
be the square of the distance between vortex $i$ and vortex $j$. With this notation, inserting (\ref{eq:GF4b}) into (\ref{eq:GF3}) and using (\ref{eq:alpha3}), one can derive the equations of motion for the $i$th vortex to be
\begin{eqnarray}
\dot{x}_i & = & -\Psi (\alpha) \sum_{i<j} \Gamma_j \frac{y_i - y_j}{r^{4-\alpha}_{ij}}~, \label{eq:PV1} \\
\dot{y}_i & = & \Psi (\alpha) \sum_{i<j} \Gamma_j \frac{x_i - x_j}{r^{4-\alpha}_{ij}}~, \label{eq:PV2}
\end{eqnarray}
for $\alpha\neq 2$. When $\alpha = 2$, inserting (\ref{eq:GF4a}) into (\ref{eq:GF3}) and again using (\ref{eq:alpha3}) yields
\begin{eqnarray}
\dot{x}_i & = & -\frac{ 1 }{2 \pi} \sum_{i<j} \Gamma_j \frac{y_i - y_j}{r^2_{ij}}~, \label{eq:PV1-2} \\
\dot{y}_i & = & \frac{ 1 }{2 \pi} \sum_{i<j} \Gamma_j \frac{x_i - x_j}{r^2_{ij}}~. \label{eq:PV2-2}
\end{eqnarray}

\subsection{Invariant quantities and integrability of the 3 vortex problem}

With these equations of motion it is possible to prove 
the following 
\begin{proposition}
The three-vortex problem for the $\alpha$-model possesses the conserved quantities
\begin{eqnarray}
H_\alpha&=&-\Psi(\alpha)\sum_{i<j}\frac{\Gamma_i\Gamma_j}{r_{ij}^{2-\alpha}} \qquad \alpha \neq 2~, \label{eq:Halpha} \\
Q&=&\sum_{i} \Gamma_i x_i~, \label{eq:QP1} \\
P&=&\sum_i\Gamma_i y_i~, \label{eq:QP2} \\
I&=&\sum_i \Gamma_i (x_i^2+y_i^2)~. \label{eq:I}
\end{eqnarray}
For the case $\alpha = 2$ one has the known Hamiltonian
\begin{equation}
H_2 = -\frac{1}{2 \pi}\sum_{i<j}\Gamma_i\Gamma_j \ln r_{ij}~. 
\label{eq:H2} 
\end{equation}
\end{proposition}
\emph{Proof}. 
First recognise that these quantities represent respectively the Hamiltonian, the two components of the linear momentum and the angular impulse. Only the Hamiltonian depends on $\alpha$ and the passage from $\alpha < 2$ to $\alpha > 2$ shows a fundamental change in the topology of the system.  
Define the canonical Poisson bracket for the N-vortex problem
\begin{equation}
[f,g]_P=\sum_i \frac{1}{\Gamma_i}\left(\frac{\partial f}{\partial x_i}\frac{\partial g}{\partial y_i}-\frac{\partial g}{\partial x_i}\frac{\partial f}{\partial y_i}\right)~.
\label{eq:Poisson}
\end{equation}
Making use of it, the proof of the invariance of the Hamiltonian simply comes from the observation that $\dot{H}_{\alpha} = [H_\alpha , H_\alpha]_P =0$. The conservation of $Q,P$ and $I$ can be proved instead from simple substitution and noting that $[Q , H_\alpha]_P = [P , H_\alpha]_P = [I , H_\alpha]_P = 0$. 

By Liouville's theorem, the previous proposition implies that the 3-vortex problem for the $\alpha$-models is completely integrable. 

The system also has an invariance for scaling, as stated in the following
\begin{proposition}
The equations of motion (\ref{eq:PV1})-(\ref{eq:PV2}) are invariant with respect to the transformations
\begin{eqnarray}
(x,y) & \rightarrow & \lambda (x',y')~, \\
t & \rightarrow & \lambda^{4-\alpha} t'~,
\end{eqnarray}
where $\lambda \in {\textbf R}$.
\end{proposition}
Proof comes from direct substitution.

Lastly, the system has two invariants linked to the circulation:
\begin{proposition}
The equations of motion (\ref{eq:PV1})-(\ref{eq:PV2}) yield the conservation of the total circulation 
\begin{equation}
\gamma = \sum_i \Gamma_i~, 
\label{eq:little gamma}
\end{equation}
and of the angular momentum 
\begin{eqnarray}
V_{\alpha} & = & \sum_i \Gamma_i \textbf{k} \cdot \left( \textbf{r}_i \times \dot{\textbf{r}}_i \right) \nonumber \\
&=& \Psi(\alpha)  \sum_{i<j} \frac{\Gamma_i \Gamma_j}{r_{ij}^{2 - \alpha}}  ~.
\label{eq:Valpha}
\end{eqnarray}
\end{proposition}
\emph{Proof}.
The conservation of (\ref{eq:little gamma}) comes directly from Kelvin's circulation theorem and, together with the conservation of $P, \, Q$ yields the conservation of the centre of circulation 
\begin{equation}
\textbf{C} = \frac{\sum_i \Gamma_i \textbf{r}_i }{ \sum_i \Gamma_i}~. 
\label{eq:centre circulation}
\end{equation}
In the case of $\alpha = 2$, the angular momentum is the virial 
\begin{equation}
V_2 = \frac{1 }{ 2 \pi} \sum_{i<j} \Gamma_i \Gamma_j ~, 
\label{eq:virial}
\end{equation}
which is a conserved scalar. For $\alpha \neq 2$ one has the interesting result $V_{\alpha} = - H_{\alpha}$, which implies $[V_{\alpha} , H_\alpha]_P = - [H_\alpha , H_\alpha]_P =  0$. 

The proof of invariance of the previous quantities can also be demonstrated a priori. It is easy to check that the Lagrangian function \cite{chapman1978ideal,badin2018variational} 
\begin{equation}
L_\alpha = H_\alpha + V_\alpha~,
\label{eq:Lalpha}
\end{equation}
here generalized for the $\alpha$-models, satisfies corresponding continuous symmetries for time and space translations, rotations and scaling. Application of Noether's theorem \cite{chapman1978ideal,badin2018variational} yields thus the conservation of $H_\alpha, \, Q, \, P, \, I$ and $V_\alpha$ respectively.


 
\subsection{Equations of relative motion}

In the following, it will be 
convenient to have the equations of motion written in terms of the mutual squared distances between the point vortices, also known as equations of relative motion \cite{groblispezielle,novikov1975dynamics}. 
Using the canonical Poisson bracket (\ref{eq:Poisson}), one has
\begin{eqnarray}
\frac{d {r}^2_{ij}}{dt}  & = &  \left[ {r}^2_{ij}, H_\alpha \right]_P \nonumber \\
&=& 2 (\alpha - 2 ) \Psi ( \alpha ) \sum_{k\neq i\neq j}\Gamma_k A \sigma \left( \frac{1}{r_{jk}^{4-\alpha}} - \frac{1}{r_{ki}^{4-\alpha}} \right) ~. 
\label{eq:r2dt}
\end{eqnarray}
Notice that for $\alpha = 2$ we have
\begin{eqnarray}
\frac{d {r}^2_{ij}}{dt}  & = &  \left[ {r}^2_{ij}, H_2 \right]_P \nonumber \\
& = &  \frac{2}{\pi}\sum_{k\neq i\neq j} \Gamma_k A \sigma \left( \frac{1}{r_{jk}^{2}} - \frac{1}{r_{ki}^{2}} \right)~. \label{eq:r2dt-2} 
\end{eqnarray}
In (\ref{eq:r2dt})-(\ref{eq:r2dt-2}), the quantity $A$ 
is the area of the triangle with $r_{ij}, r_{jk}, r_{ki}$ as side-lengths and $\sigma$ is the orientation of the same triangle, so that $\sigma=1$ if the vortices are arranged counter-clockwise and $\sigma=-1$ otherwise. Without loss of generality, in the following we will set $\sigma=1$. 

Alternatively, one has
\begin{equation} 
\dot{r}_{ij}  =  (\alpha - 2 ) \Psi ( \alpha ) \sum_{k\neq i\neq j}\frac{\Gamma_k}{ \rho} \left( \frac{r_{ki}}{r_{jk}^{3-\alpha}} - \frac{r_{jk}}{r_{ki}^{3-\alpha}} \right)~, \label{eq:EM1}
\end{equation}
where $\rho = r_{ij}  r_{jk}  r_{ki} / A$.
For $\alpha = 2$ 
\begin{equation}
\dot{r}_{ij}  =   \sum_{k\neq i\neq j}\frac{\Gamma_k}{ \pi \rho} \left( \frac{r_{ki}}{r_{jk}} - \frac{r_{jk}}{r_{ki}} \right) ~. 
\label{eq:EM2a} 
\end{equation}

\begin{remark}
From (\ref{eq:EM1})-(\ref{eq:EM2a}) we see that all equilibria and relative equilibria of the three-vortex problem for the $\alpha$-models correspond to cases in which either 
\begin{equation}
r_{ij}=r_{jk}=r_{ki}~, 
\label{eq: condition equilateral}
\end{equation}
or 
\begin{equation}
A=0~, 
\label{eq: condition collinear}
\end{equation}
and hence the vortices must form an equilateral triangle or collinear configuration. 
This also holds and is well-known for the classical three-vortex problem ($\alpha = 2$). In particular, since the vector field  (\ref{eq:EM1})-(\ref{eq:EM2a}) is smooth near equilateral triangle configurations, except at the point of collapse, it is easy to see that (\ref{eq: condition equilateral}) is also a sufficient condition for equilibrium \cite{hernandez2007collisions}.
For the collinear configuration, the condition (\ref{eq: condition collinear}) gives rise to three possibilities, corresponding to (i) an absolute equilibrium; (ii) a relative equilibrium, i.e. a rigidly translating or rotating collinear configuration of vortices; or (iii) an evolution of the collinear configuration to a triangle with nonzero area and then, due to the symmetries of the system, to a commuted collinear configuration \cite{hernandez2007collisions}. In order for the configuration to remain collinear (and hence to avoid case (iii)), the system should also satisfy the condition
\begin{equation}
\dot{A}=0~. 
\label{eq: condition collinear 2}
\end{equation}
Finally, the system does not allow for partial collapse, i.e. for collapse of two of the three vortices. The proof proceeds in complete analogy with the case $\alpha = 2$ \cite{hernandez2007collisions}, and will not be reported here. 
\label{remark1}
\end{remark}

It should be noted that in (\ref{eq:Halpha})-(\ref{eq:I}), only the Hamiltonian is written as a function of the relative distances. In terms of these, it is possible to prove the conservation of the quantity
\begin{equation}
M = \frac{1}{2} \sum_{i<j} \Gamma_i \Gamma_j r_{ij}^2~,
\label{eq:M}
\end{equation}
which is the squared relative angular momentum with respect to the centre of circulation. This follows from the observation that $[H_\alpha , M]_P=0$. Notice that $M$ does not depend on $\alpha$.

\subsection{Nambu formulation of the equations of motion for three point vortices}

Using the conserved quantities $H_\alpha$ and $M$,  one can rewrite the equations of relative motion for three point vortices as
\begin{equation}
\dot{r}_{ij} = \frac{1}{\Gamma_1 \Gamma_2 \Gamma_3 \rho} \left( \frac{\partial M}{\partial r_{jk}} \frac{\partial H_\alpha}{\partial r_{ki}} - \frac{\partial M}{\partial r_{ki}} \frac{\partial H_\alpha}{\partial r_{jk}}\right)~,
\label{eq:EM3}
\end{equation}
or, in vector notation,
\begin{equation}
\dot{{\textbf r}} =  \frac{1}{\Gamma_1 \Gamma_2 \Gamma_3 \rho} \nabla M \times \nabla H_\alpha~.
\label{eq:EM4}
\end{equation}
Equation (\ref{eq:EM4}) is the non-canonical Nambu representation for the point-vortex problem. Notice that, as in underlying partial differential equations (\ref{eq:36}), the Nambu representation depends on $\alpha$ through $H_\alpha$ and not through the coefficients of the equations. 

Following the arguments reported in the Introduction, (\ref{eq:EM4}) shows that the motion can be geometrically identified along the manifold given by the intersection of the level sets of $H_\alpha$ and $M$. In the following we will use this approach to identify the necessary conditions for collapse.
 
\section{Collapse for the $\alpha$-models}

\subsection{Collapse manifold}

Define
\begin{eqnarray}
B_\alpha & = & - \frac{H_\alpha}{\Psi ( \alpha )  \Gamma_1 \Gamma_2 \Gamma_3}~, \label{eq:balpha} \\
C & = &  \frac{2 M}{\Gamma_1 \Gamma_2 \Gamma_3}~, \label{eq:c} 
\end{eqnarray} 
with the special case
\begin{equation}
B_2  =  - \frac{2 \pi H_2}{  \Gamma_1 \Gamma_2 \Gamma_3}~. \label{eq:b2} 
\end{equation}
With these definitions, Equations (\ref{eq:Halpha}), (\ref{eq:M}), and (\ref{eq:H2}),  become
\begin{eqnarray}
B_\alpha &=& \frac{r_{12}^{\alpha - 2}}{\Gamma_3}+\frac{r_{23}^{\alpha - 2}}{\Gamma_1}+\frac{r_{31}^{\alpha - 2}}{\Gamma_2}~, \label{eq:balpha_eq} \\
C & = & \frac{r^2_{12}}{\Gamma_3}+\frac{r^2_{23}}{\Gamma_1}+\frac{r^2_{31}}{\Gamma_2}~, \label{eq:M_eq} 
\end{eqnarray}
with the special case
\begin{equation}
B_2 = \frac{\ln r_{12}}{\Gamma_3}+ \frac{\ln r_{23}}{\Gamma_1}+ \frac{\ln r_{31}}{\Gamma_2}~. \label{eq:b2_eq} 
\end{equation}
The state of total collapse corresponds to $r_{12}=r_{23}=r_{31}=0$, and so we must require
\begin{equation}
C = 0~.
\label{eq:c=0}
\end{equation}

This immediately implies that one of the circulations carries a sign opposite the other two.  
We can also use (\ref{eq:M_eq}) to express one of the variables, e.g. $r_{31}$, as a function of the other variables. Since $r_{ij} \ge 0$, it follows that
\begin{equation}
r_{31} = \sqrt{-\Gamma_2 \left( \frac{r^2_{12}}{\Gamma_3} + \frac{r^2_{23}}{\Gamma_1} \right)} ~,
\label{eq:r12c}
\end{equation}
which must hold for all values of $\alpha$. Equation (\ref{eq:balpha_eq}) yields
\begin{equation}
r_{31} = \left[ -\Gamma_2 \left( \frac{r^{\alpha - 2}_{12}}{\Gamma_3} + \frac{r^{\alpha - 2}_{23}}{\Gamma_1} - B_\alpha \right) \right]^{\frac{1}{\alpha - 2}} ~, \qquad \alpha \neq 2~.
\label{eq:r12alpha}
\end{equation}
The manifolds containing candidate trajectories for collapse are thus given by the intersections of the surfaces defined by (\ref{eq:r12c}) and (\ref{eq:r12alpha}). 

\subsection{Admissible cone}

In spite of the above, the condition that trajectories are constrained to the intersection of the surfaces is not sufficient for collapse.  This is a consequence of the ambiguity arising from the mutual distances coordinate space. Consider the following geometric argument (see e.g. \cite{aref2007point}). Given the triangular inequalities
\begin{equation}
 r_{ij}  \le  r_{jk} + r_{ki}~,  
 \label{eq:triangular ineq} 
\end{equation}
one has
\begin{eqnarray}
& \left( r_{12} + r_{23} + r_{31}  \right)\left( r_{12} + r_{23} - r_{31}  \right) \nonumber \\
& \times \left( r_{12} - r_{23} + r_{31}  \right)\left( - r_{12} + r_{23} + r_{31}  \right) \ge 0~.
 \label{eq:triangular ineq2} 
\end{eqnarray}

Geometrically, the set of points 
inside the region
satisfying (\ref{eq:triangular ineq2}) describe a conic section. Only the trajectories that lie inside this so-called \emph{admissible cone} represent realizable vortex triangles. It is trivial to see that the roots of (\ref{eq:triangular ineq2}) for e.g. $r_{31}$, defining the boundaries of the admissible cone, are $r_{31}= \pm (r_{12} + r_{23})$ and $r_{31}=\pm ( r_{12} - r_{23})$. Notice that, by definition, we are interested in the regions where $r_{ij} \ge 0$. 

\begin{remark}
In the classical point vortex model ($\alpha=2$), collapse is self-similar and can occur for any value of the Hamiltonian, but has a strict relationship on the vortex circulations, i.e. \cite{aref1979motion,novikov1979vortex}
\begin{equation}
\sum_i 1 / \Gamma_i=0~.
\label{eq:cond self sim alpha 2}
\end{equation}
This condition arises from special properties of the logarithm. Notice also that the same condition implies $V_2 = 0$. 
For the symmetric case $\Gamma_1=\Gamma_2=1$, $\Gamma_3=-\Gamma<0$ that will be considered later, collapse can thus only occur for $\Gamma=1 / 2$.    
\end{remark}

If initial conditions are chosen to lie on the collapse manifold and within the admissible cone, then the evolution of the orbit moves toward the collapse point in either forward or backward time, provided the manifold does not contain any equilibria or relative equilibria.  Since all equilibria and relative equilibria are either collinear or equilateral, it is easy to check the existence of such trajectories. In what follows we will see that families of collapsing and expanding configurations come in pairs, in analogy with results from the classical Euler three-vortex problem.  In the next Section, the collapse manifold and the admissible cone will be studied for the special case of SQG, i.e. for $\alpha = 1$. 
 
\section{Collapse for the SQG equations}

\subsection{Equations of motion} \label{sec:sqg}

For the SQG equations, i.e. for $\alpha = 1$, one has
\begin{equation}
\Psi (1 ) = - \frac{ 1 }{2 \pi}~,
\label{eq:GF6}
\end{equation}
and thus
\begin{equation}
G^{(1)} ({\textbf r} ) = - \frac{ 1 }{2 \pi {\textbf r}}~,
\label{eq:GF7}
\end{equation}
which is the well-known Green's function discussed in, e.g., \cite{heldetal95}. 

With this, the equations of motion become
\begin{eqnarray}
\dot{x}_i & = & \frac{ 1 }{2 \pi} \sum_{i<j} \Gamma_j \frac{y_i - y_j}{r^{3}_{ij}}~, \label{eq:PV1-sqg} \\
\dot{y}_i & = & -\frac{ 1 }{2 \pi} \sum_{i<j} \Gamma_j \frac{x_i - x_j}{r^{3}_{ij}}~. \label{eq:PV2-sqg}
\end{eqnarray}
The Hamiltonian is
\begin{equation}
H_1 = \frac{1}{2 \pi}\sum_{i<j}\frac{\Gamma_i\Gamma_j}{r_{ij}}~.
\label{eq:H1}
\end{equation}
In the following, the subscript $1$ will be omitted.
Using the results for the $\alpha$-models, one arrives at the equations of relative motion for three point vortices,
\begin{eqnarray}
\frac{d {r}^2_{ij}}{dt}  & = & \left[ {r}^2_{ij}, H \right]_P \nonumber \\
& = &  \frac{1}{\pi} \Gamma_k A \sigma \left( \frac{1}{r_{jk}^{3}} - \frac{1}{r_{ki}^{3}} \right) ~, 
\label{eq:r2dt-sqg} 
\end{eqnarray}
or, equivalently,
\begin{equation}
\dot{r}_{ij} =  \frac{\Gamma_k}{ 2 \pi \rho} \left( \frac{r_{ki}}{r_{jk}^{2}} - \frac{r_{jk}}{r_{ki}^{2}} \right)~.
\label{eq:EM2}
\end{equation}
 
\subsection{Necessary conditions for collapse}

For three SQG point vortices, one has
\begin{equation}
H=\frac{1}{2\pi}\left(\frac{\Gamma_1\Gamma_2}{r_{12}}+\frac{\Gamma_2\Gamma_3}{r_{23}}+\frac{\Gamma_3\Gamma_1}{r_{31}}\right)~,
\end{equation}
and therefore
\begin{equation}
B = \frac{1}{\Gamma_3 r_{12}}+\frac{1}{\Gamma_1 r_{23}}+\frac{1}{\Gamma_2 r_{31}} ~. 
\label{eq:B-sqg}
\end{equation}
We recall that $B$ was defined as
\begin{equation}
B= \frac{2\pi H}{\Gamma_1 \Gamma_2 \Gamma_3}~.
\end{equation}
Note
that Equation (\ref{eq:B-sqg}) does not allow for partial collapse, i.e. the
instance of two vortices coalescing while the third remains separate.
In particular, if any one of the mutual distances $r_{ij}$ tends to
zero, then another must also tend to zero in order for $B$ to remain
constant.  However, the latter is a total collapse scenario.

To allow for direct comparison with the case $\alpha=2$, for which (\ref{eq:cond self sim alpha 2}) must hold for collapse, we choose $\Gamma_1=\Gamma_3=1$, $\Gamma_2=-\Gamma<0$. This is the simplest situation that can lead to examples of both self-similar and non-self-similar collapse. We refer to this as the \emph{symmetric} case. 

For simplicity and ease of notation, we set also $x=r_{12}$, $y=r_{23}$, $z=r_{31}$, where $x,y,z \in \textbf{R}^+$.  


Equations (\ref{eq:B-sqg}) and (\ref{eq:c=0}) become 
\begin{eqnarray}
B&=&\frac{1}{x}+\frac{1}{y}-\frac{1}{\Gamma z}~,\label{eq:H-sqg} \\
0&=&x^2+y^2- \frac{ z^2}{\Gamma}~,\label{eq:M-sqg}
\end{eqnarray}
respectively. 

Notice that in this notation the roots of (\ref{eq:triangular ineq2}) satisfying $(x,y,z) \ge 0$ are given by $z=x+y$ and $z=\pm(x-y)$.

\begin{figure}
 \includegraphics[scale=0.65]{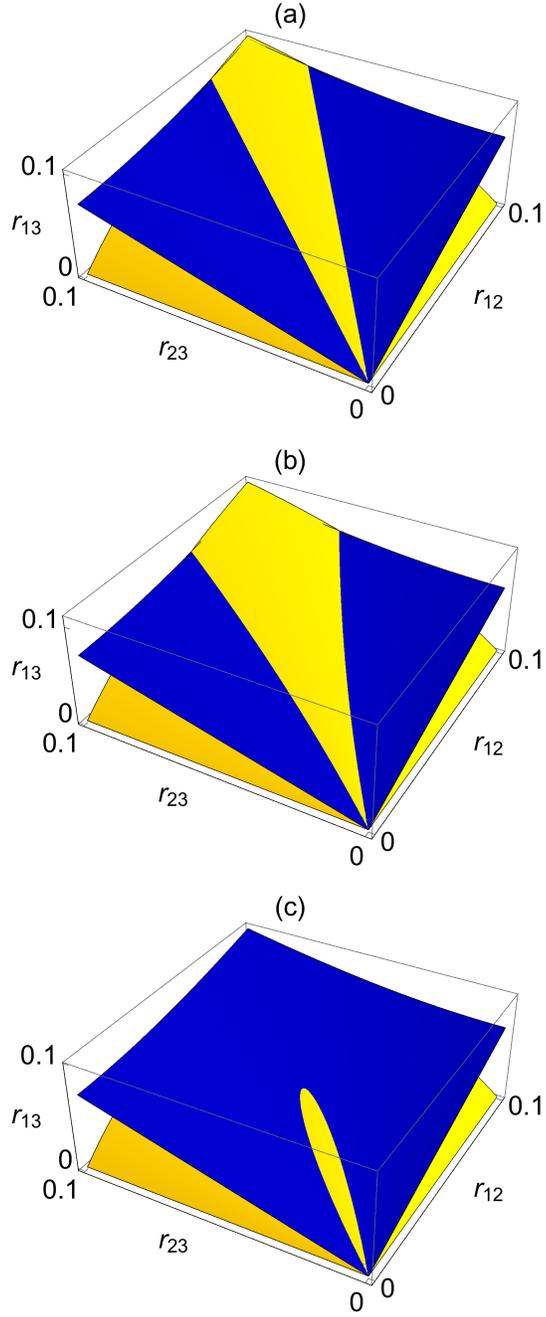}%
 \caption{\label{fig1} Surfaces given by the equations (\ref{eq:r12c}) (blue) and (\ref{eq:r12alpha}) (yellow) for the choice of values $\Gamma_1=\Gamma_3=1$, $\Gamma_2=-0.49$ and (a) $H=0$, (b) $H=-0.1$ and (c) $H=0.1$ (bottom).}
 \end{figure}
  
We now state the following
\begin{proposition}
The SQG three-vortex model possesses collapsing solutions.  For the symmetric case, collapse can be self-similar or not. When $H=0$, collapse is self-similar, while for $H \neq 0$, collapse must be non-self-similar.
\end{proposition}
The proof of the proposition will be demonstrated over the next two subsections, and numerical examples will then be reported.

\subsubsection{Self-similar collapse}

Consider first the case where the side-lengths of the triangle have a linear scaling, i.e.
\begin{eqnarray}
\frac{y}{x} &=& k_1~, \\
\frac{z}{x} &=& k_2~,
\end{eqnarray}
with $k_1, k_2 \in {\textbf R}^+$.  
The replacement of these in (\ref{eq:H-sqg}) yields
\begin{equation}
B=\left(1+\frac{1}{k_1}-\frac{1}{\Gamma k_2}\right)\frac{1}{x} ~.
\end{equation}
In order for the above equation to hold for all $x>0$, we must have
\begin{eqnarray}
1+\frac{1}{k_1}-\frac{1}{\Gamma k_2} &=& 0~, \label{eq:simcond}
\end{eqnarray}
and hence
\begin{eqnarray}
B &=& 0~. \label{eq:b=0} 
\end{eqnarray}
Pairing this with (\ref{eq:M-sqg}) gives a relationship between $k_1$ and $k_2$, that is
\begin{equation}
k_2^3=\frac{k_1(1+k_1^2)}{k_1+1}~.
\end{equation}

\begin{remark}
It should be emphasized that (\ref{eq:b=0}) implies that the linear scaling for SQG-collapse requires $H=0$. This differs strikingly from the classical point vortex model ($\alpha=2$), where self-similar collapse requires (\ref{eq:cond self sim alpha 2}). No analogous property arises in the SQG case nor for the generic $\alpha$-models. The same condition implies $V_2 = 0$, which does not arise in the SQG case nor for the generic $\alpha$-models, see (\ref{eq:Valpha}).    
\end{remark}

Indeed,  we will see that any choice of $\Gamma$ within a certain interval can lead to collapse with linear scaling for SQG when $H=0$.
 
\begin{lemma}
Any choice of $\Gamma$, with $\Gamma_*<\Gamma<0.5$, where $\Gamma_*\approx 0.387464$, can lead to self-similar collapse for SQG when $H=0$.
\end{lemma}
  
\emph{Proof}. Consider the planar level sets $z=Z$ for fixed values $Z \in \textbf{R}^+$. We are interested in the intersections of these curves for small values of $Z$.  Equations (\ref{eq:H-sqg})-(\ref{eq:M-sqg}) are
\begin{eqnarray}
\frac{1}{\Gamma Z}&=&\frac{1}{x}+\frac{1}{y}~, \label{eq:hyperbolic} \\
\frac{ Z^2}{\Gamma}&=&x^2+y^2~. \label{eq:quadratic}
\end{eqnarray}
There are three possible scenarios for the intersections: \emph{(i)} the two curves do not intersect; \emph{(ii)} they intersect along the line $y=x$; \emph{(iii)} they intersect along two different lines, symmetrically placed about $y=x$, one of them representing collapsing solutions while the other consists of expanding solutions.  
In order to guarantee the existence of the 
intersections, we require that the quadratic curve (\ref{eq:quadratic}) lies ``above'' the hyperbolic curve (\ref{eq:hyperbolic}) with respect to the line $y=x$.  Along that line, 
\begin{eqnarray}
\frac{1}{\Gamma Z} &=& \frac{2}{x}~, \\
\frac{ Z^2 }{\Gamma} &=& 2x^2~.
\end{eqnarray}
Thus, let 
\begin{eqnarray}
x_H &=& 2 \Gamma Z~, \\
x_M &=& \sqrt{\frac{1}{2 \Gamma}}Z~.
\end{eqnarray}
The condition of one curve lying above the other is $ x_M > x_H$. It is evident that this is true provided $0 < \Gamma < 1 / 2$.  Notice that when $\Gamma=1 / 2$ one has $x_M=x_H$ and the level curves meet along the line of equilateral triangles.

To complete the proof, one should prove that the trajectories always lie within the admissible cone. In the region $(x,y,z) \ge 0$ and with the assumption $0<\Gamma<0.5$, the plane $z=x+y$ never meets the surface defined by equation (\ref{eq:H-sqg}) since, in the defined region, $(x+y)^2>x^2+y^2> \Gamma(x^2+y^2)$.  Thus the curves of intersection never meet the $z=x+y$ boundary of the admissible cone. 

Inserting $z=y-x$ into (\ref{eq:H-sqg}) and (\ref{eq:M-sqg}) yield the linear relationships
\begin{equation}
y=\frac{(1+\sqrt{(2-\Gamma)\Gamma})}{1-\Gamma}x,
\end{equation}
and
\begin{equation}
y=\frac{(1+\sqrt{1+4\Gamma^2})}{2\Gamma}x,
\end{equation}
respectively.
The two lines coincide for $\Gamma=\Gamma_*\approx 0.387464$, so that, i.e. for $\Gamma=\Gamma_*$ the manifold given by the intersection of $H$ and $M$ coincides with the boundary of the admissible cone.
 
Finally, because of symmetry, $z=x-y$ will give the same value of $\Gamma_*$.  Thus collapse occurs only for the interval $\Gamma_*<\Gamma<0.5$.
 
 \begin{figure}
 \includegraphics[scale=0.65]{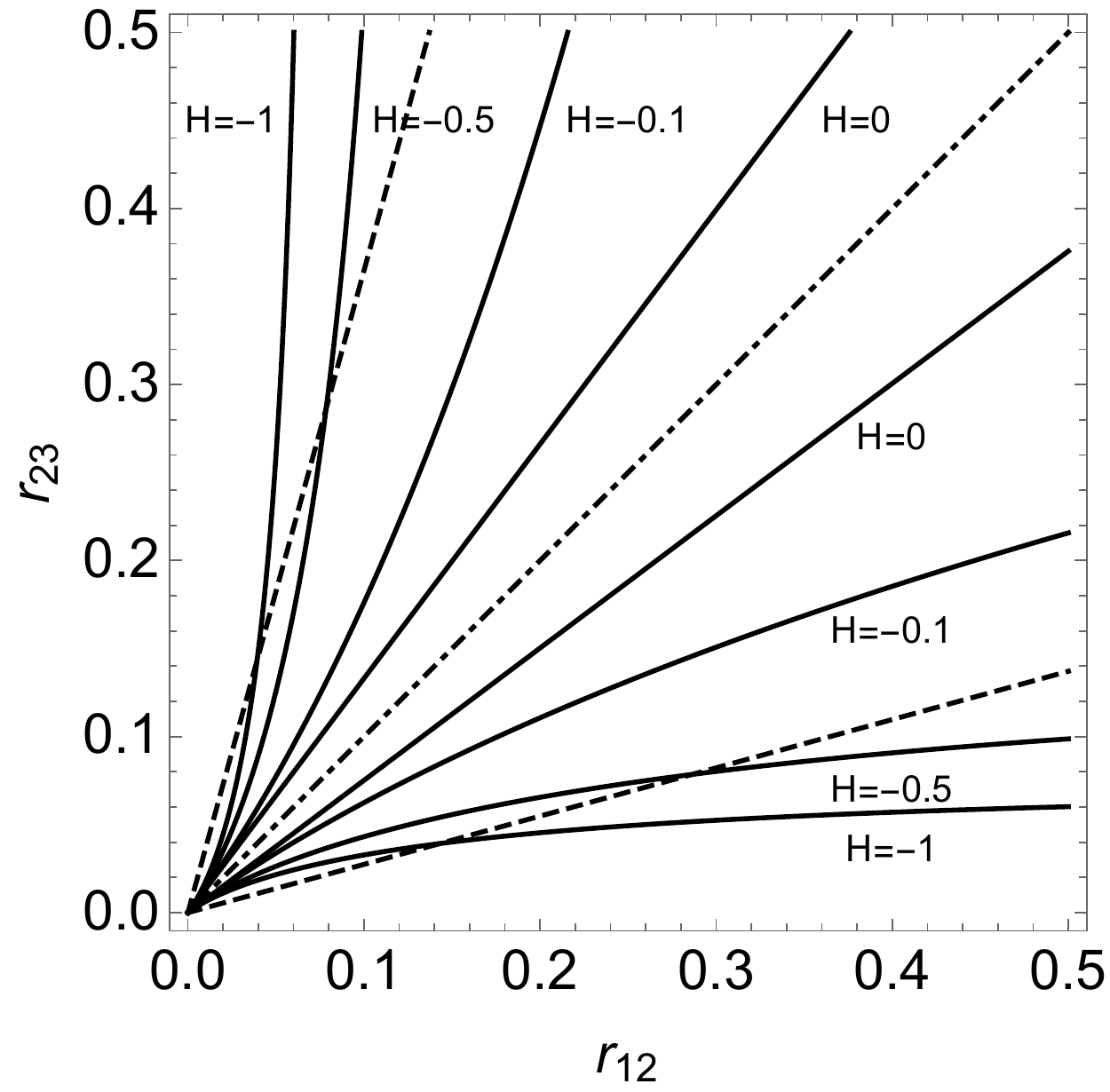}%
 \caption{\label{fig2} Intersections of the curves given by the equations (\ref{eq:r12alpha}) and (\ref{eq:r12c}) for the choice of values $\Gamma_1=\Gamma_3=1$, $\Gamma_2=-0.49$ and $H$ ranging from $H = -1$ (most outer curves) to $H = 0$ (most inner curves). The dashed lines represent the boundary of the admissible cone. The dot-dashed line represents the line of equilateral triangles defined by $r_{12}=r_{23}$.}
 \end{figure}
 
\subsubsection{Non self-similar collapse}

Based on the above arguments, we know that the linear scaling assumption cannot possibly lead to collapse for $H \neq 0$ (i.e. $B\neq 0$). We would now like to pose the question as to whether there can be collapse for non-zero energy levels. Using the geometric approach, we work with level curves as in the previous section.  The equations with $B\neq 0$ are now
\begin{eqnarray}
\frac{1}{\Gamma} \left(\frac{1}{Z}+B\right)&=& \frac{1}{x}+\frac{1}{y}~, \label{eq:hyperbolic-nonlinear} \\
\frac{1}{\Gamma} Z^2&=& x^2+y^2~. \label{eq:quadratic-nonlinear}
\end{eqnarray}
As before, in the the region of interest $x>0, y>0$ the first curve is concave up and the second is concave down.  Further, the first curve only lies in this region of the plane when $\frac{1}{\Gamma} \left(\frac{1}{Z}+B\right)>0$, which implies that we must request $\frac{1}{Z}+B>0$.  This will clearly hold when $Z$ is sufficiently small, which is in agreement with the fact that for collapse we are interested in the level curves close to the origin, i.e. set by $Z\rightarrow 0$.

We wish to identify where the quadratic curve lies ``above'' the hyperbolic curve along the line $y=x$ as $x,y\rightarrow 0$.  This will guarantee the two curves of intersections as in the earlier section.  We have
\begin{eqnarray}
x_H&=&\frac{2 \Gamma}{ \left(\frac{1}{Z}+B\right)}~, \\
x_M&=&\sqrt{\frac{1}{2 \Gamma}}Z~.
\end{eqnarray}
Thus, $x_M>x_H$ holds in the limit $Z\rightarrow 0$ only if $0 < \Gamma < 1 / 2$ exactly as in the $H=0$ case. For $H \neq 0$, it is not possible to evaluate analytically the lower bound for $\Gamma$ from the constraint given by the admissible cone. In the next section, this will thus be calculated numerically for specific values of $H$. 
 \begin{figure}
 \includegraphics[scale=0.65]{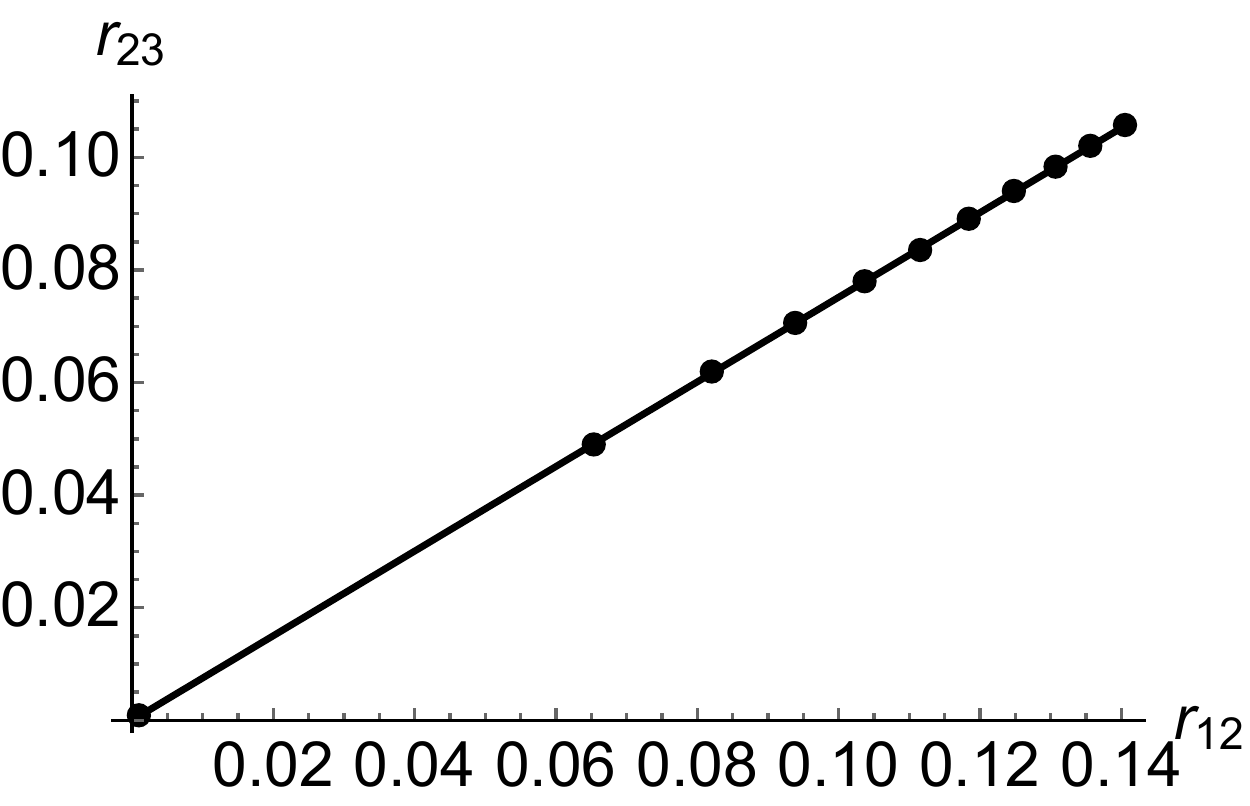}%
 \caption{\label{fig3} Black dots: data from numerical integration of the equations of motion (\ref{eq:EM2}) when $\Gamma_1=\Gamma_3=1$, $\Gamma_2=-0.49$.  Time increases from right to left and the initial conditions were chosen to satisfy (\ref{eq:H-sqg})-(\ref{eq:M-sqg}) with $B=H=0$.
   \, Solid line: analytical solutions of (\ref{eq:H-sqg})-(\ref{eq:M-sqg}), with $B=0$, given by $r_{23}=0.751484 \, r_{12}$.}
 \end{figure}

\subsection{Examples of SQG collapse}
 
In the previous subsection we have identified candidate curves for collapse.
In what follows, we give examples of collapse solutions for particular choices of the parameters. Figure \ref{fig1} shows the surfaces given by the equations (\ref{eq:r12c}) (blue) and (\ref{eq:r12alpha}) (yellow) for three different values of $H$, corresponding to $H=0$ (top panel), $H=-0.1$ (middle panel) and $H=0.1$ (bottom panel), and $\Gamma_1=\Gamma_3=1$, $|\Gamma_*|<|\Gamma_2|<0.5$. In particular, results are shown for $\Gamma_2=-0.49$. As predicted, the surfaces intersect along two curves passing through the origin and are symmetric about the line $r_{12}=r_{23}$, but have substantial differences for different values of $H$. 



 \begin{figure}
 \includegraphics[scale=0.65]{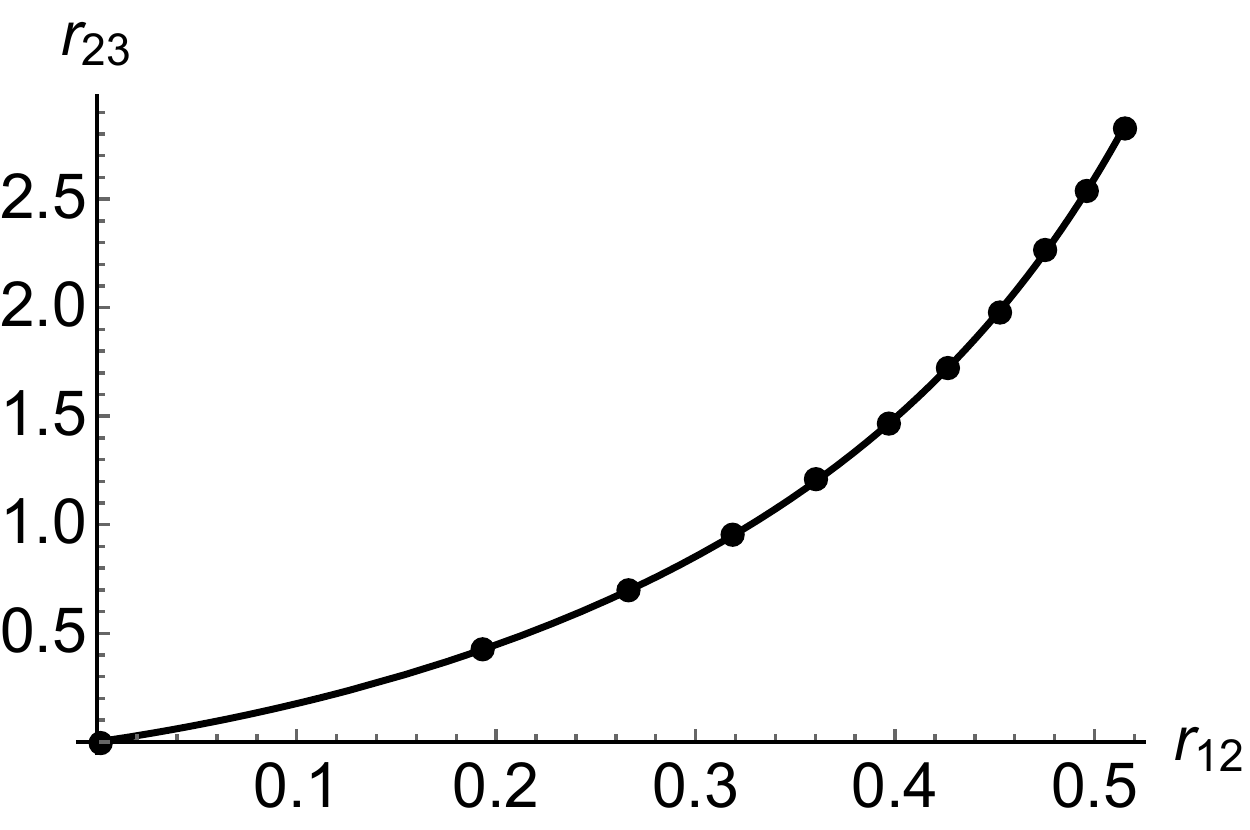}%
 \caption{\label{fig4} Black dots: data from numerical integration of the equations of motion (\ref{eq:EM2}) when $\Gamma_1=\Gamma_3=1$, $\Gamma_2=-0.49$.  Time increases from right to left and the initial conditions were chosen to satisfy (\ref{eq:H-sqg})-(\ref{eq:M-sqg}) with $H=-0.1$. 
 Solid curve: analytical solutions of (\ref{eq:H-sqg})-(\ref{eq:M-sqg}) with the specified choice of parameters.}
 \end{figure}
 
\subsubsection{Examples of SQG collapse for $H = 0$}

The intersection of the surfaces given by the equations (\ref{eq:r12c}) (blue) and (\ref{eq:r12alpha}) (yellow)  for $H=0$ are shown in the top panel of Figure \ref{fig1}. The corresponding projections on the $(r_{12} , r_{23})$-plane are shown in Figure \ref{fig2}. For $H = 0$, the intersections are straight lines (seen as the two symmetric lines closer to the $r_{12}=r_{23}$ line, marked as dot-dashed) contained within the admissible cone. 

Solving equations (\ref{eq:H-sqg})-(\ref{eq:M-sqg}) for $z$ yields a quartic equation in $x$ and $y$. If $H = 0$, the solutions are linear.  There are two solutions that meet the region $x,y>0$, namely
 \begin{eqnarray}
y=0.751484 x~, \label{eq:solution linear 1}\\
y=1.3307 x~. \label{eq:solution linear 2}
\end{eqnarray}
%

Using these two sets of equations allows us to generate initial conditions that may lead to collapse when integrating the system (\ref{eq:EM2}).  Of course, collapse corresponds to a singularity of the system, and the equations blow up as this point is approached. 
Therefore, in order to verify that the simulation corresponds to a collapse trajectory, we compare it with the analytical solution provided by Equations (\ref{eq:solution linear 1})-(\ref{eq:solution linear 2}). 



Our results show that Equation (\ref{eq:solution linear 1}) corresponds to collapse, while Equation (\ref{eq:solution linear 2}) corresponds to expansion (collapse in backward time). The data from the simulation is shown using black dots in Figure \ref{fig3}, while the solid line shows the analytical solution satisfying (\ref{eq:solution linear 1}).
%
 
\subsubsection{Examples of SQG collapse for $H \ne 0$}

\paragraph{ $H<0$.} The intersection of the surfaces given by the equations (\ref{eq:r12c}) (blue) and (\ref{eq:r12alpha}) (yellow)  for $H<0$ are shown in the middle panel of Figure \ref{fig1}. The corresponding projections on the $(r_{12} , r_{23})$-plane are shown in Figure \ref{fig2}. For different values of $H < 0$, the intersections are seen as symmetric lines with nonlinear profiles, more and more distant from the $r_{12}=r_{23}$ line as the value of $H$ diminishes, and with non-zero intersections with the boundaries of the admissible cone. 

In this case, we set $H = -0.1$ and again generate initial conditions using Equations (\ref{eq:H-sqg})-(\ref{eq:M-sqg}). Figure \ref{fig4} shows the time integration of (\ref{eq:EM2}) using these initial conditions. The trajectory follows an evidently nonlinear path in the projection onto the ($r_{12}$,$r_{23}$)-plane.  Data from the simulation are shown with black dots, and the projection of the relevant intersection of the surfaces generated by (\ref{eq:H-sqg})-(\ref{eq:M-sqg}) onto the plane is shown as solid curve. 

\begin{figure}
 \includegraphics[scale=0.65]{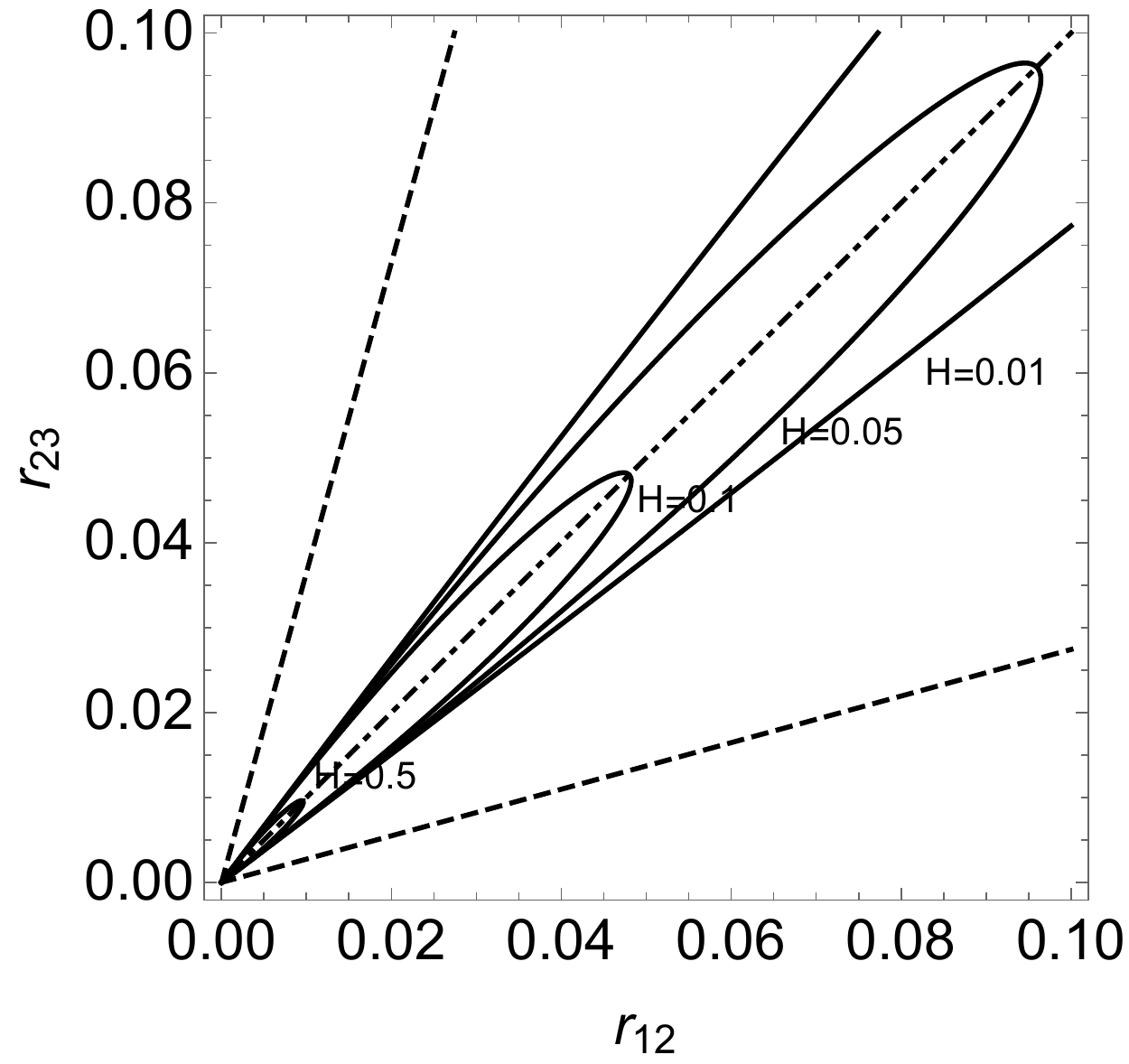}%
 \caption{\label{fig5} Intersections of the curves given by the equations (\ref{eq:r12alpha}) and (\ref{eq:r12c}) for the choice of values $\Gamma_1=\Gamma_3=1$, $\Gamma_2=-0.49$ and $H$ ranging from $H = 0.01$ (most outer curves) to $H = 0.5$ (most inner curves). The dashed lines represent the boundary of the admissible cone. The dot-dashed line represents the line of equilateral triangles defined by $r_{12}=r_{23}$.}
 \end{figure}

\paragraph{ $H>0$.} The intersection of the surfaces given by the equations (\ref{eq:r12c}) (blue) and (\ref{eq:r12alpha}) (yellow)  for $H>0$ are shown in the lower panel of Figure \ref{fig1}. Figure \ref{fig5} shows the intersections of the curves given by the equations (\ref{eq:r12alpha}) and (\ref{eq:r12c}) for the choice of values $\Gamma_1=\Gamma_3=1$, $\Gamma_2=-0.49$ and $H$ ranging from $H = 0.01$ (most outer curves) to $H = 0.5$ (most inner curves). The dashed lines represent the boundary of the admissible cone. The dot-dashed line represent the line given by the equilateral triangles defined by $r_{12}=r_{23}$. 

Also in this case, it is clear that trajectories follow a non self-similar evolution. The results for these values of $H$ are however very different than the cases with $H \le 0$. The trajectories in the $(r_{12},r_{23})$-phase plane appear as homoclinic orbits connected to the origin. The connection to homoclinic orbits is however only apparent. Consider in fact the part of a trajectory given by a fixed value of $H$. If the initial condition corresponds to a forward time integration toward the origin, the necessary conditions for collapse are satisfied. However, if the initial conditions correspond to a forward time integration away the origin, the trajectory will approach the line $r_{12}=r_{23}$, i.e. the trajectory approaches an equilateral triangle relative equilibrium.  In other words, each intersection contains two heteroclinic orbits: one that approaches collapse in backward time and an equilateral triangle in forward time, and one that does the opposite.  Therefore, there are no infinitely expanding solutions in this scenario.  This case is particularly striking because there is no analogous scenario in the classical Euler point vortex problem.

\section{Summary and discussion}
%
In this article we have presented point-vortex equations for the $\alpha$-models, giving particular attention to the SQG case. Derivation of the invariant quantities shows a dependence on the parameter $\alpha$ only in the Hamiltonian, a result that was already found for the infinite-dimensional case \cite{blender2015hydrodynamic}. The point vortex equations were expressed via Nambu dynamics, which allowed for the use of not only the Hamiltonian but also of the Casimir of the system. As a result, trajectories were found to follow the intersection of fixed level sets of the Hamiltonian and Casimir, as already studied by \cite{Makhaldiani07,Makhaldiani12,MuellerNevir14}. 

Within this setting, we studied the collapse of solutions for the three point-vortex model. Results show that for SQG the collapse can be either self-similar or less. Self-similarity is restricted to a zero value of the Hamiltonian, while non-self-similarity appears for non-zero values of the same. For both cases, collapse is allowed for any choice of the circulations within a permitted interval. As remarked, these results differ strikingly from the classical point vortex model, where collapse is self-similar for any value of the Hamiltonian, but has a strict relationship on the vortex circulations. 

Information on the behaviour of the SQG equations near collapse is particularly interesting, at least at a qualitative level, as it is still unknown whether the partial differential equations of the SQG model can produce singularities. As suggested in \cite{novikov1979vortex,vosbeek1997collapse}, careful examination of the initial conditions leading to collapse might be used to construct initial distributions of the active scalar for the SQG model, which might be treated as candidates for the formation of singularities. Or, vice-versa, prominent candidates of initial conditions that are thought to lead to singularities of the SQG equations, such as the one proposed by \cite{constantinetal94}, can be used to construct initial configurations of point vortices. The eventual collapse of these initial conditions can thus give \emph{qualitative} information about the formation of singularities. It is also interesting to note that while collapse can happen in both self-similar and non-self-similar ways, the partial differential equations of SQG are thought to give raise to singularities only in a self-similar way.   

As stated in the Introduction, the $\alpha$-models can be used to give a representation of balanced models of geophysical flows, such as the atmosphere and the ocean. The Hamiltonian nature of these models is valid at scales between the forcing and dissipation scales. It should be underlined that due to the extremely high values of the Reynolds' number associated to geophysical flows, this condition is satisfied for a large class of atmospheric and oceanic dynamics.  
The $\alpha$-models here analysed can however be used to represent also higher-order balanced models of geophysical flows, such as the surface semi-geostrophic (SSG) model \citep{badin2013surface,ragone2016study}. As semi-geostrophic dynamics is known for the formation of singularities, which represent the finite-time formation of fronts, the analysis of the collapse of the corresponding point-vortex models could give interesting qualitative information for the relation between collapse and singularities.

\begin{acknowledgments}
The initial ideas of this work were developed during two visits of AMB to the University of Hamburg, financed by the Cluster of Excellence ``Integrated Climate System Analysis and Prediction'' (CliSAP). We would like to thank two anonymous referees for suggestions that helped to improve the manuscript. 
GB was partially funded by the DFG research grants TRR181, 1740, BA 5068/8-1 and BA 5068/9-1.
AMB wishes to thank Claire Postlethwaite
for useful discussions related to this topic. 

\end{acknowledgments}

\bibliography{references}

\begin{thebibliography}{104}%
\makeatletter
\providecommand \@ifxundefined [1]{%
 \@ifx{#1\undefined}
}%
\providecommand \@ifnum [1]{%
 \ifnum #1\expandafter \@firstoftwo
 \else \expandafter \@secondoftwo
 \fi
}%
\providecommand \@ifx [1]{%
 \ifx #1\expandafter \@firstoftwo
 \else \expandafter \@secondoftwo
 \fi
}%
\providecommand \natexlab [1]{#1}%
\providecommand \enquote  [1]{``#1''}%
\providecommand \bibnamefont  [1]{#1}%
\providecommand \bibfnamefont [1]{#1}%
\providecommand \citenamefont [1]{#1}%
\providecommand \href@noop [0]{\@secondoftwo}%
\providecommand \href [0]{\begingroup \@sanitize@url \@href}%
\providecommand \@href[1]{\@@startlink{#1}\@@href}%
\providecommand \@@href[1]{\endgroup#1\@@endlink}%
\providecommand \@sanitize@url [0]{\catcode `\\12\catcode `\$12\catcode
  `\&12\catcode `\#12\catcode `\^12\catcode `\_12\catcode `\%12\relax}%
\providecommand \@@startlink[1]{}%
\providecommand \@@endlink[0]{}%
\providecommand \url  [0]{\begingroup\@sanitize@url \@url }%
\providecommand \@url [1]{\endgroup\@href {#1}{\urlprefix }}%
\providecommand \urlprefix  [0]{URL }%
\providecommand \Eprint [0]{\href }%
\providecommand \doibase [0]{http://dx.doi.org/}%
\providecommand \selectlanguage [0]{\@gobble}%
\providecommand \bibinfo  [0]{\@secondoftwo}%
\providecommand \bibfield  [0]{\@secondoftwo}%
\providecommand \translation [1]{[#1]}%
\providecommand \BibitemOpen [0]{}%
\providecommand \bibitemStop [0]{}%
\providecommand \bibitemNoStop [0]{.\EOS\space}%
\providecommand \EOS [0]{\spacefactor3000\relax}%
\providecommand \BibitemShut  [1]{\csname bibitem#1\endcsname}%
\let\auto@bib@innerbib\@empty
\bibitem [{\citenamefont {Blumen}(1978)}]{blumen78}%
  \BibitemOpen
  \bibfield  {author} {\bibinfo {author} {\bibfnamefont {W.}~\bibnamefont
  {Blumen}},\ }\href@noop {} {\bibfield  {journal} {\bibinfo  {journal} {J.
  Atmos. Sci.}\ }\textbf {\bibinfo {volume} {35}},\ \bibinfo {pages} {774}
  (\bibinfo {year} {1978})}\BibitemShut {NoStop}%
\bibitem [{\citenamefont {Holm}\ \emph {et~al.}(1998)\citenamefont {Holm},
  \citenamefont {Marsden},\ and\ \citenamefont {Ratiu}}]{holm1998euler}%
  \BibitemOpen
  \bibfield  {author} {\bibinfo {author} {\bibfnamefont {D.}~\bibnamefont
  {Holm}}, \bibinfo {author} {\bibfnamefont {J.}~\bibnamefont {Marsden}}, \
  and\ \bibinfo {author} {\bibfnamefont {T.}~\bibnamefont {Ratiu}},\
  }\href@noop {} {\bibfield  {journal} {\bibinfo  {journal} {Advances in
  Mathematics}\ }\textbf {\bibinfo {volume} {137}},\ \bibinfo {pages} {1}
  (\bibinfo {year} {1998})}\BibitemShut {NoStop}%
\bibitem [{\citenamefont {Pierrehumbert}\ \emph {et~al.}(1994)\citenamefont
  {Pierrehumbert}, \citenamefont {Held},\ and\ \citenamefont
  {Swanson}}]{pierrehumbertetal94}%
  \BibitemOpen
  \bibfield  {author} {\bibinfo {author} {\bibfnamefont {R.}~\bibnamefont
  {Pierrehumbert}}, \bibinfo {author} {\bibfnamefont {I.}~\bibnamefont {Held}},
  \ and\ \bibinfo {author} {\bibfnamefont {K.}~\bibnamefont {Swanson}},\
  }\href@noop {} {\bibfield  {journal} {\bibinfo  {journal} {Chaos Solit.
  Fract.}\ }\textbf {\bibinfo {volume} {4}},\ \bibinfo {pages} {1111} (\bibinfo
  {year} {1994})}\BibitemShut {NoStop}%
\bibitem [{\citenamefont {Smith}\ \emph {et~al.}(2002)\citenamefont {Smith},
  \citenamefont {Boccaletti}, \citenamefont {Henning}, \citenamefont {Marinov},
  \citenamefont {Tam}, \citenamefont {Held},\ and\ \citenamefont
  {Vallis}}]{smith2002turbulent}%
  \BibitemOpen
  \bibfield  {author} {\bibinfo {author} {\bibfnamefont {K.~S.}\ \bibnamefont
  {Smith}}, \bibinfo {author} {\bibfnamefont {G.}~\bibnamefont {Boccaletti}},
  \bibinfo {author} {\bibfnamefont {C.}~\bibnamefont {Henning}}, \bibinfo
  {author} {\bibfnamefont {I.}~\bibnamefont {Marinov}}, \bibinfo {author}
  {\bibfnamefont {C.}~\bibnamefont {Tam}}, \bibinfo {author} {\bibfnamefont
  {I.}~\bibnamefont {Held}}, \ and\ \bibinfo {author} {\bibfnamefont {G.~K.}\
  \bibnamefont {Vallis}},\ }\href@noop {} {\bibfield  {journal} {\bibinfo
  {journal} {J. Fluid Mech.}\ }\textbf {\bibinfo {volume} {469}},\ \bibinfo
  {pages} {13} (\bibinfo {year} {2002})}\BibitemShut {NoStop}%
\bibitem [{\citenamefont {Tran}\ and\ \citenamefont
  {Shepherd}(2002)}]{tran2002constraints}%
  \BibitemOpen
  \bibfield  {author} {\bibinfo {author} {\bibfnamefont {C.}~\bibnamefont
  {Tran}}\ and\ \bibinfo {author} {\bibfnamefont {T.}~\bibnamefont
  {Shepherd}},\ }\href@noop {} {\bibfield  {journal} {\bibinfo  {journal}
  {Physica D}\ }\textbf {\bibinfo {volume} {165}},\ \bibinfo {pages} {199}
  (\bibinfo {year} {2002})}\BibitemShut {NoStop}%
\bibitem [{\citenamefont {Tran}(2004)}]{tran2004nonlinear}%
  \BibitemOpen
  \bibfield  {author} {\bibinfo {author} {\bibfnamefont {C.}~\bibnamefont
  {Tran}},\ }\href@noop {} {\bibfield  {journal} {\bibinfo  {journal} {Physica
  D}\ }\textbf {\bibinfo {volume} {191}},\ \bibinfo {pages} {137} (\bibinfo
  {year} {2004})}\BibitemShut {NoStop}%
\bibitem [{\citenamefont {Tran}\ \emph {et~al.}(2010)\citenamefont {Tran},
  \citenamefont {Dritschel},\ and\ \citenamefont {Scott}}]{tran2010effective}%
  \BibitemOpen
  \bibfield  {author} {\bibinfo {author} {\bibfnamefont {C.}~\bibnamefont
  {Tran}}, \bibinfo {author} {\bibfnamefont {D.}~\bibnamefont {Dritschel}}, \
  and\ \bibinfo {author} {\bibfnamefont {R.}~\bibnamefont {Scott}},\
  }\href@noop {} {\bibfield  {journal} {\bibinfo  {journal} {Physical Review
  E}\ }\textbf {\bibinfo {volume} {81}},\ \bibinfo {pages} {016301} (\bibinfo
  {year} {2010})}\BibitemShut {NoStop}%
\bibitem [{\citenamefont {Schorghofer}(2000)}]{schorghofer2000universality}%
  \BibitemOpen
  \bibfield  {author} {\bibinfo {author} {\bibfnamefont {N.}~\bibnamefont
  {Schorghofer}},\ }\href@noop {} {\bibfield  {journal} {\bibinfo  {journal}
  {Phys. Rev. E}\ }\textbf {\bibinfo {volume} {61}},\ \bibinfo {pages} {6568}
  (\bibinfo {year} {2000})}\BibitemShut {NoStop}%
\bibitem [{\citenamefont {Burgess}\ and\ \citenamefont
  {Shepherd}(2013)}]{burgess2013spectral}%
  \BibitemOpen
  \bibfield  {author} {\bibinfo {author} {\bibfnamefont {B.}~\bibnamefont
  {Burgess}}\ and\ \bibinfo {author} {\bibfnamefont {T.}~\bibnamefont
  {Shepherd}},\ }\href@noop {} {\bibfield  {journal} {\bibinfo  {journal} {J.
  Fluid Mech.}\ }\textbf {\bibinfo {volume} {725}},\ \bibinfo {pages} {332}
  (\bibinfo {year} {2013})}\BibitemShut {NoStop}%
\bibitem [{\citenamefont {Burgess}\ \emph {et~al.}(2015)\citenamefont
  {Burgess}, \citenamefont {Scott},\ and\ \citenamefont
  {Shepherd}}]{burgess2015kraichnan}%
  \BibitemOpen
  \bibfield  {author} {\bibinfo {author} {\bibfnamefont {B.}~\bibnamefont
  {Burgess}}, \bibinfo {author} {\bibfnamefont {R.}~\bibnamefont {Scott}}, \
  and\ \bibinfo {author} {\bibfnamefont {T.}~\bibnamefont {Shepherd}},\
  }\href@noop {} {\bibfield  {journal} {\bibinfo  {journal} {J. Fluid Mech.}\
  }\textbf {\bibinfo {volume} {767}},\ \bibinfo {pages} {467} (\bibinfo {year}
  {2015})}\BibitemShut {NoStop}%
\bibitem [{\citenamefont {Venaille}\ \emph {et~al.}(2015)\citenamefont
  {Venaille}, \citenamefont {Dauxois},\ and\ \citenamefont
  {Ruffo}}]{venaille2015violent}%
  \BibitemOpen
  \bibfield  {author} {\bibinfo {author} {\bibfnamefont {A.}~\bibnamefont
  {Venaille}}, \bibinfo {author} {\bibfnamefont {T.}~\bibnamefont {Dauxois}}, \
  and\ \bibinfo {author} {\bibfnamefont {S.}~\bibnamefont {Ruffo}},\
  }\href@noop {} {\bibfield  {journal} {\bibinfo  {journal} {Phys. Rev. E}\
  }\textbf {\bibinfo {volume} {92}},\ \bibinfo {pages} {011001} (\bibinfo
  {year} {2015})}\BibitemShut {NoStop}%
\bibitem [{\citenamefont {Foussard}\ \emph {et~al.}(2017)\citenamefont
  {Foussard}, \citenamefont {Berti}, \citenamefont {Perrot},\ and\
  \citenamefont {Lapeyre}}]{foussard2017relative}%
  \BibitemOpen
  \bibfield  {author} {\bibinfo {author} {\bibfnamefont {A.}~\bibnamefont
  {Foussard}}, \bibinfo {author} {\bibfnamefont {S.}~\bibnamefont {Berti}},
  \bibinfo {author} {\bibfnamefont {X.}~\bibnamefont {Perrot}}, \ and\ \bibinfo
  {author} {\bibfnamefont {G.}~\bibnamefont {Lapeyre}},\ }\href@noop {}
  {\bibfield  {journal} {\bibinfo  {journal} {J. Fluid Mech.}\ }\textbf
  {\bibinfo {volume} {821}},\ \bibinfo {pages} {358} (\bibinfo {year}
  {2017})}\BibitemShut {NoStop}%
\bibitem [{\citenamefont {Conte}\ and\ \citenamefont
  {Badin}(2018)}]{conti18arxiv}%
  \BibitemOpen
  \bibfield  {author} {\bibinfo {author} {\bibfnamefont {G.}~\bibnamefont
  {Conte}}\ and\ \bibinfo {author} {\bibfnamefont {G.}~\bibnamefont {Badin}},\
  }\href@noop {} {\bibfield  {journal} {\bibinfo  {journal} {arXiv:1805.01210}\
  } (\bibinfo {year} {2018})}\BibitemShut {NoStop}%
\bibitem [{\citenamefont {Held}\ \emph {et~al.}(1995)\citenamefont {Held},
  \citenamefont {Pierrehumbert}, \citenamefont {Garner},\ and\ \citenamefont
  {Swanson}}]{heldetal95}%
  \BibitemOpen
  \bibfield  {author} {\bibinfo {author} {\bibfnamefont {I.~M.}\ \bibnamefont
  {Held}}, \bibinfo {author} {\bibfnamefont {R.~T.}\ \bibnamefont
  {Pierrehumbert}}, \bibinfo {author} {\bibfnamefont {S.~T.}\ \bibnamefont
  {Garner}}, \ and\ \bibinfo {author} {\bibfnamefont {K.~L.}\ \bibnamefont
  {Swanson}},\ }\href@noop {} {\bibfield  {journal} {\bibinfo  {journal} {J.
  Fluid Mech.}\ }\textbf {\bibinfo {volume} {282}},\ \bibinfo {pages} {1}
  (\bibinfo {year} {1995})}\BibitemShut {NoStop}%
\bibitem [{\citenamefont {Lapeyre}(2017)}]{lapeyre17}%
  \BibitemOpen
  \bibfield  {author} {\bibinfo {author} {\bibfnamefont {G.}~\bibnamefont
  {Lapeyre}},\ }\href@noop {} {\bibfield  {journal} {\bibinfo  {journal}
  {Fluids}\ }\textbf {\bibinfo {volume} {2}},\ \bibinfo {pages} {7} (\bibinfo
  {year} {2017})}\BibitemShut {NoStop}%
\bibitem [{\citenamefont {Badin}\ and\ \citenamefont
  {Crisciani}(2018)}]{badin2018variational}%
  \BibitemOpen
  \bibfield  {author} {\bibinfo {author} {\bibfnamefont {G.}~\bibnamefont
  {Badin}}\ and\ \bibinfo {author} {\bibfnamefont {F.}~\bibnamefont
  {Crisciani}},\ }\href@noop {} {\emph {\bibinfo {title} {Variational
  Formulation of Fluid and Geophysical Fluid Dynamics: Mechanics, Symmetries
  and Conservation Laws}}}\ (\bibinfo  {publisher} {Springer},\ \bibinfo {year}
  {2018})\BibitemShut {NoStop}%
\bibitem [{\citenamefont {Tulloch}\ and\ \citenamefont
  {Smith}(2006)}]{tulloch2006theory}%
  \BibitemOpen
  \bibfield  {author} {\bibinfo {author} {\bibfnamefont {R.}~\bibnamefont
  {Tulloch}}\ and\ \bibinfo {author} {\bibfnamefont {K.}~\bibnamefont
  {Smith}},\ }\href@noop {} {\bibfield  {journal} {\bibinfo  {journal} {Proc.
  Natl. Acad. Sci. U.S.A.}\ }\textbf {\bibinfo {volume} {103}},\ \bibinfo
  {pages} {14690} (\bibinfo {year} {2006})}\BibitemShut {NoStop}%
\bibitem [{\citenamefont {Capet}\ \emph {et~al.}(2008)\citenamefont {Capet},
  \citenamefont {Klein}, \citenamefont {Hua}, \citenamefont {Lapeyre},\ and\
  \citenamefont {Mc{W}illiams}}]{capet2008surface}%
  \BibitemOpen
  \bibfield  {author} {\bibinfo {author} {\bibfnamefont {X.}~\bibnamefont
  {Capet}}, \bibinfo {author} {\bibfnamefont {P.}~\bibnamefont {Klein}},
  \bibinfo {author} {\bibfnamefont {B.}~\bibnamefont {Hua}}, \bibinfo {author}
  {\bibfnamefont {G.}~\bibnamefont {Lapeyre}}, \ and\ \bibinfo {author}
  {\bibfnamefont {J.}~\bibnamefont {Mc{W}illiams}},\ }\href@noop {} {\bibfield
  {journal} {\bibinfo  {journal} {J. Fluid Mech.}\ }\textbf {\bibinfo {volume}
  {604}},\ \bibinfo {pages} {165} (\bibinfo {year} {2008})}\BibitemShut
  {NoStop}%
\bibitem [{\citenamefont {Scott}(2006)}]{scott2006local}%
  \BibitemOpen
  \bibfield  {author} {\bibinfo {author} {\bibfnamefont {R.}~\bibnamefont
  {Scott}},\ }\href@noop {} {\bibfield  {journal} {\bibinfo  {journal} {Phys.
  Fluids}\ }\textbf {\bibinfo {volume} {18}},\ \bibinfo {pages} {116601}
  (\bibinfo {year} {2006})}\BibitemShut {NoStop}%
\bibitem [{\citenamefont {Mc{W}illiams}(2016)}]{McWilliams2016}%
  \BibitemOpen
  \bibfield  {author} {\bibinfo {author} {\bibfnamefont {J.~C.}\ \bibnamefont
  {Mc{W}illiams}},\ }\href@noop {} {\bibfield  {journal} {\bibinfo  {journal}
  {Proc. Roy. Soc. A}\ }\textbf {\bibinfo {volume} {472}} (\bibinfo {year}
  {2016})}\BibitemShut {NoStop}%
\bibitem [{\citenamefont {Badin}\ \emph {et~al.}(2011)\citenamefont {Badin},
  \citenamefont {Tandon},\ and\ \citenamefont {Mahadevan}}]{badin2011lateral}%
  \BibitemOpen
  \bibfield  {author} {\bibinfo {author} {\bibfnamefont {G.}~\bibnamefont
  {Badin}}, \bibinfo {author} {\bibfnamefont {A.}~\bibnamefont {Tandon}}, \
  and\ \bibinfo {author} {\bibfnamefont {A.}~\bibnamefont {Mahadevan}},\
  }\href@noop {} {\bibfield  {journal} {\bibinfo  {journal} {J. Phys.
  Oceanogr.}\ }\textbf {\bibinfo {volume} {41}},\ \bibinfo {pages} {2080}
  (\bibinfo {year} {2011})}\BibitemShut {NoStop}%
\bibitem [{\citenamefont {Shcherbina}\ \emph {et~al.}(2015)\citenamefont
  {Shcherbina}, \citenamefont {Sundermeyer}, \citenamefont {Kunze},
  \citenamefont {D'Asaro}, \citenamefont {Badin}, \citenamefont {Birch},
  \citenamefont {Brunner-Suzuki}, \citenamefont {Callies}, \citenamefont
  {Cervantes}, \citenamefont {Claret}, \citenamefont {Concannon}, \citenamefont
  {Early}, \citenamefont {Ferrari}, \citenamefont {Goodman}, \citenamefont
  {Harcourt}, \citenamefont {Klymak}, \citenamefont {Lee}, \citenamefont
  {Lelong}, \citenamefont {Levine}, \citenamefont {Lien}, \citenamefont
  {Mahadevan}, \citenamefont {Mc{W}illiams}, \citenamefont {Molemaker},
  \citenamefont {Mukherjee}, \citenamefont {Nash}, \citenamefont
  {{\"O}zg{\"o}kmen}, \citenamefont {Pierce}, \citenamefont {Ramachandran},
  \citenamefont {Samelson}, \citenamefont {Sanford}, \citenamefont {Shearman},
  \citenamefont {Skyllingstad}, \citenamefont {Smith}, \citenamefont {Tandon},
  \citenamefont {Taylor}, \citenamefont {Terray}, \citenamefont {Thomas},\ and\
  \citenamefont {Ledwell}}]{Shcherbina2015}%
  \BibitemOpen
  \bibfield  {author} {\bibinfo {author} {\bibfnamefont {A.~Y.}\ \bibnamefont
  {Shcherbina}}, \bibinfo {author} {\bibfnamefont {M.~A.}\ \bibnamefont
  {Sundermeyer}}, \bibinfo {author} {\bibfnamefont {E.}~\bibnamefont {Kunze}},
  \bibinfo {author} {\bibfnamefont {E.}~\bibnamefont {D'Asaro}}, \bibinfo
  {author} {\bibfnamefont {G.}~\bibnamefont {Badin}}, \bibinfo {author}
  {\bibfnamefont {D.}~\bibnamefont {Birch}}, \bibinfo {author} {\bibfnamefont
  {A.-M. E.~G.}\ \bibnamefont {Brunner-Suzuki}}, \bibinfo {author}
  {\bibfnamefont {J.}~\bibnamefont {Callies}}, \bibinfo {author} {\bibfnamefont
  {B.~T.~K.}\ \bibnamefont {Cervantes}}, \bibinfo {author} {\bibfnamefont
  {M.}~\bibnamefont {Claret}}, \bibinfo {author} {\bibfnamefont
  {B.}~\bibnamefont {Concannon}}, \bibinfo {author} {\bibfnamefont
  {J.}~\bibnamefont {Early}}, \bibinfo {author} {\bibfnamefont
  {R.}~\bibnamefont {Ferrari}}, \bibinfo {author} {\bibfnamefont
  {L.}~\bibnamefont {Goodman}}, \bibinfo {author} {\bibfnamefont {R.~R.}\
  \bibnamefont {Harcourt}}, \bibinfo {author} {\bibfnamefont {J.~M.}\
  \bibnamefont {Klymak}}, \bibinfo {author} {\bibfnamefont {C.~M.}\
  \bibnamefont {Lee}}, \bibinfo {author} {\bibfnamefont {M.-P.}\ \bibnamefont
  {Lelong}}, \bibinfo {author} {\bibfnamefont {M.~D.}\ \bibnamefont {Levine}},
  \bibinfo {author} {\bibfnamefont {R.-C.}\ \bibnamefont {Lien}}, \bibinfo
  {author} {\bibfnamefont {A.}~\bibnamefont {Mahadevan}}, \bibinfo {author}
  {\bibfnamefont {J.~C.}\ \bibnamefont {Mc{W}illiams}}, \bibinfo {author}
  {\bibfnamefont {M.~J.}\ \bibnamefont {Molemaker}}, \bibinfo {author}
  {\bibfnamefont {S.}~\bibnamefont {Mukherjee}}, \bibinfo {author}
  {\bibfnamefont {J.~D.}\ \bibnamefont {Nash}}, \bibinfo {author}
  {\bibfnamefont {T.}~\bibnamefont {{\"O}zg{\"o}kmen}}, \bibinfo {author}
  {\bibfnamefont {S.~D.}\ \bibnamefont {Pierce}}, \bibinfo {author}
  {\bibfnamefont {S.}~\bibnamefont {Ramachandran}}, \bibinfo {author}
  {\bibfnamefont {R.~M.}\ \bibnamefont {Samelson}}, \bibinfo {author}
  {\bibfnamefont {T.~B.}\ \bibnamefont {Sanford}}, \bibinfo {author}
  {\bibfnamefont {R.~K.}\ \bibnamefont {Shearman}}, \bibinfo {author}
  {\bibfnamefont {E.~D.}\ \bibnamefont {Skyllingstad}}, \bibinfo {author}
  {\bibfnamefont {K.~S.}\ \bibnamefont {Smith}}, \bibinfo {author}
  {\bibfnamefont {A.}~\bibnamefont {Tandon}}, \bibinfo {author} {\bibfnamefont
  {J.~R.}\ \bibnamefont {Taylor}}, \bibinfo {author} {\bibfnamefont {E.~A.}\
  \bibnamefont {Terray}}, \bibinfo {author} {\bibfnamefont {L.~N.}\
  \bibnamefont {Thomas}}, \ and\ \bibinfo {author} {\bibfnamefont {J.~R.}\
  \bibnamefont {Ledwell}},\ }\href@noop {} {\bibfield  {journal} {\bibinfo
  {journal} {BAMS}\ }\textbf {\bibinfo {volume} {96}},\ \bibinfo {pages} {1257}
  (\bibinfo {year} {2015})}\BibitemShut {NoStop}%
\bibitem [{\citenamefont {Mukiibi}\ \emph {et~al.}(2016)\citenamefont
  {Mukiibi}, \citenamefont {Badin},\ and\ \citenamefont
  {Serra}}]{mukiibi2016three}%
  \BibitemOpen
  \bibfield  {author} {\bibinfo {author} {\bibfnamefont {D.}~\bibnamefont
  {Mukiibi}}, \bibinfo {author} {\bibfnamefont {G.}~\bibnamefont {Badin}}, \
  and\ \bibinfo {author} {\bibfnamefont {N.}~\bibnamefont {Serra}},\
  }\href@noop {} {\bibfield  {journal} {\bibinfo  {journal} {J.Phys.
  Oceanogr.}\ }\textbf {\bibinfo {volume} {46}},\ \bibinfo {pages} {1509}
  (\bibinfo {year} {2016})}\BibitemShut {NoStop}%
\bibitem [{\citenamefont {Badin}(2014)}]{badin2014role}%
  \BibitemOpen
  \bibfield  {author} {\bibinfo {author} {\bibfnamefont {G.}~\bibnamefont
  {Badin}},\ }\href@noop {} {\bibfield  {journal} {\bibinfo  {journal} {Phys.
  Fluids}\ }\textbf {\bibinfo {volume} {26}},\ \bibinfo {pages} {096603}
  (\bibinfo {year} {2014})}\BibitemShut {NoStop}%
\bibitem [{\citenamefont {Carton}(2009)}]{carton2009instability}%
  \BibitemOpen
  \bibfield  {author} {\bibinfo {author} {\bibfnamefont {X.}~\bibnamefont
  {Carton}},\ }\href@noop {} {\bibfield  {journal} {\bibinfo  {journal} {J.
  Atmos. Sci.}\ }\textbf {\bibinfo {volume} {66}},\ \bibinfo {pages} {1051}
  (\bibinfo {year} {2009})}\BibitemShut {NoStop}%
\bibitem [{\citenamefont {Dritschel}(2011)}]{dritschel2011exact}%
  \BibitemOpen
  \bibfield  {author} {\bibinfo {author} {\bibfnamefont {D.}~\bibnamefont
  {Dritschel}},\ }\href@noop {} {\bibfield  {journal} {\bibinfo  {journal}
  {Geophys. Astro. Fluid Dyn.}\ }\textbf {\bibinfo {volume} {105}},\ \bibinfo
  {pages} {368} (\bibinfo {year} {2011})}\BibitemShut {NoStop}%
\bibitem [{\citenamefont {Harvey}\ and\ \citenamefont
  {Ambaum}(2011)}]{harvey2011perturbed}%
  \BibitemOpen
  \bibfield  {author} {\bibinfo {author} {\bibfnamefont {B.}~\bibnamefont
  {Harvey}}\ and\ \bibinfo {author} {\bibfnamefont {M.}~\bibnamefont
  {Ambaum}},\ }\href@noop {} {\bibfield  {journal} {\bibinfo  {journal}
  {Geophys. Astro. Fluid Dyn.}\ }\textbf {\bibinfo {volume} {105}},\ \bibinfo
  {pages} {377} (\bibinfo {year} {2011})}\BibitemShut {NoStop}%
\bibitem [{\citenamefont {Harvey}\ \emph {et~al.}(2011)\citenamefont {Harvey},
  \citenamefont {Ambaum},\ and\ \citenamefont
  {Carton}}]{harvey2011instability}%
  \BibitemOpen
  \bibfield  {author} {\bibinfo {author} {\bibfnamefont {B.}~\bibnamefont
  {Harvey}}, \bibinfo {author} {\bibfnamefont {M.}~\bibnamefont {Ambaum}}, \
  and\ \bibinfo {author} {\bibfnamefont {X.}~\bibnamefont {Carton}},\
  }\href@noop {} {\bibfield  {journal} {\bibinfo  {journal} {J. Atmos. Sci.}\
  }\textbf {\bibinfo {volume} {68}},\ \bibinfo {pages} {964} (\bibinfo {year}
  {2011})}\BibitemShut {NoStop}%
\bibitem [{\citenamefont {Bembenek}\ \emph {et~al.}(2015)\citenamefont
  {Bembenek}, \citenamefont {Poulin},\ and\ \citenamefont
  {Waite}}]{bembenek2015realizing}%
  \BibitemOpen
  \bibfield  {author} {\bibinfo {author} {\bibfnamefont {E.}~\bibnamefont
  {Bembenek}}, \bibinfo {author} {\bibfnamefont {F.}~\bibnamefont {Poulin}}, \
  and\ \bibinfo {author} {\bibfnamefont {M.}~\bibnamefont {Waite}},\
  }\href@noop {} {\bibfield  {journal} {\bibinfo  {journal} {J. Phys.
  Oceanogr.}\ }\textbf {\bibinfo {volume} {45}},\ \bibinfo {pages} {1376}
  (\bibinfo {year} {2015})}\BibitemShut {NoStop}%
\bibitem [{\citenamefont {Carton}\ \emph {et~al.}(2016)\citenamefont {Carton},
  \citenamefont {Ciani}, \citenamefont {Verron}, \citenamefont {Reinaud},\ and\
  \citenamefont {Sokolovskiy}}]{carton2016vortex}%
  \BibitemOpen
  \bibfield  {author} {\bibinfo {author} {\bibfnamefont {X.}~\bibnamefont
  {Carton}}, \bibinfo {author} {\bibfnamefont {D.}~\bibnamefont {Ciani}},
  \bibinfo {author} {\bibfnamefont {J.}~\bibnamefont {Verron}}, \bibinfo
  {author} {\bibfnamefont {J.}~\bibnamefont {Reinaud}}, \ and\ \bibinfo
  {author} {\bibfnamefont {M.}~\bibnamefont {Sokolovskiy}},\ }\href@noop {}
  {\bibfield  {journal} {\bibinfo  {journal} {Geophys. Astro. Fluid Dyn.}\
  }\textbf {\bibinfo {volume} {110}},\ \bibinfo {pages} {1} (\bibinfo {year}
  {2016})}\BibitemShut {NoStop}%
\bibitem [{\citenamefont {Badin}\ and\ \citenamefont
  {Poulin}(2018)}]{Badinpoulin2018}%
  \BibitemOpen
  \bibfield  {author} {\bibinfo {author} {\bibfnamefont {G.}~\bibnamefont
  {Badin}}\ and\ \bibinfo {author} {\bibfnamefont {F.}~\bibnamefont {Poulin}},\
  }\href@noop {} {\bibfield  {journal} {\bibinfo  {journal} {Geophys. Astro.
  Fluid Dyn.}\ } (\bibinfo {year} {2018})},\ \bibinfo {note}
  {doi:10.1080/03091929.2018.1453930}\BibitemShut {NoStop}%
\bibitem [{\citenamefont {Constantin}\ \emph
  {et~al.}(1994{\natexlab{a}})\citenamefont {Constantin}, \citenamefont
  {Majda},\ and\ \citenamefont {Tabak}}]{constantinetal94}%
  \BibitemOpen
  \bibfield  {author} {\bibinfo {author} {\bibfnamefont {P.}~\bibnamefont
  {Constantin}}, \bibinfo {author} {\bibfnamefont {A.}~\bibnamefont {Majda}}, \
  and\ \bibinfo {author} {\bibfnamefont {E.}~\bibnamefont {Tabak}},\
  }\href@noop {} {\bibfield  {journal} {\bibinfo  {journal} {Nonlinearity}\
  }\textbf {\bibinfo {volume} {7}},\ \bibinfo {pages} {1495} (\bibinfo {year}
  {1994}{\natexlab{a}})}\BibitemShut {NoStop}%
\bibitem [{\citenamefont {Constantin}\ \emph
  {et~al.}(1994{\natexlab{b}})\citenamefont {Constantin}, \citenamefont
  {Majda},\ and\ \citenamefont {Tabak}}]{constantin1994singular}%
  \BibitemOpen
  \bibfield  {author} {\bibinfo {author} {\bibfnamefont {P.}~\bibnamefont
  {Constantin}}, \bibinfo {author} {\bibfnamefont {A.}~\bibnamefont {Majda}}, \
  and\ \bibinfo {author} {\bibfnamefont {E.}~\bibnamefont {Tabak}},\
  }\href@noop {} {\bibfield  {journal} {\bibinfo  {journal} {Phys. Fluids}\
  }\textbf {\bibinfo {volume} {6}},\ \bibinfo {pages} {9} (\bibinfo {year}
  {1994}{\natexlab{b}})}\BibitemShut {NoStop}%
\bibitem [{\citenamefont {Majda}\ and\ \citenamefont
  {Tabak}(1996)}]{majda1996two}%
  \BibitemOpen
  \bibfield  {author} {\bibinfo {author} {\bibfnamefont {A.}~\bibnamefont
  {Majda}}\ and\ \bibinfo {author} {\bibfnamefont {E.}~\bibnamefont {Tabak}},\
  }\href@noop {} {\bibfield  {journal} {\bibinfo  {journal} {Physica D}\
  }\textbf {\bibinfo {volume} {98}},\ \bibinfo {pages} {515} (\bibinfo {year}
  {1996})}\BibitemShut {NoStop}%
\bibitem [{\citenamefont {Ohkitani}\ and\ \citenamefont
  {Yamada}(1997)}]{ohkitani1997inviscid}%
  \BibitemOpen
  \bibfield  {author} {\bibinfo {author} {\bibfnamefont {K.}~\bibnamefont
  {Ohkitani}}\ and\ \bibinfo {author} {\bibfnamefont {M.}~\bibnamefont
  {Yamada}},\ }\href@noop {} {\bibfield  {journal} {\bibinfo  {journal} {Phys.
  Fluids}\ }\textbf {\bibinfo {volume} {9}},\ \bibinfo {pages} {876} (\bibinfo
  {year} {1997})}\BibitemShut {NoStop}%
\bibitem [{\citenamefont {Constantin}\ \emph {et~al.}(1998)\citenamefont
  {Constantin}, \citenamefont {Nie},\ and\ \citenamefont
  {Sch{\"o}rghofer}}]{constantin1998nonsingular}%
  \BibitemOpen
  \bibfield  {author} {\bibinfo {author} {\bibfnamefont {P.}~\bibnamefont
  {Constantin}}, \bibinfo {author} {\bibfnamefont {Q.}~\bibnamefont {Nie}}, \
  and\ \bibinfo {author} {\bibfnamefont {N.}~\bibnamefont {Sch{\"o}rghofer}},\
  }\href@noop {} {\bibfield  {journal} {\bibinfo  {journal} {Phys. Lett. A}\
  }\textbf {\bibinfo {volume} {241}},\ \bibinfo {pages} {168} (\bibinfo {year}
  {1998})}\BibitemShut {NoStop}%
\bibitem [{\citenamefont {Constantin}\ and\ \citenamefont
  {Wu}(1999)}]{constantin1999behavior}%
  \BibitemOpen
  \bibfield  {author} {\bibinfo {author} {\bibfnamefont {P.}~\bibnamefont
  {Constantin}}\ and\ \bibinfo {author} {\bibfnamefont {J.}~\bibnamefont
  {Wu}},\ }\href@noop {} {\bibfield  {journal} {\bibinfo  {journal} {SIAM J.
  Math. Anal.}\ }\textbf {\bibinfo {volume} {30}},\ \bibinfo {pages} {937}
  (\bibinfo {year} {1999})}\BibitemShut {NoStop}%
\bibitem [{\citenamefont {C{\'o}rdoba}\ and\ \citenamefont
  {Fefferman}(2002{\natexlab{a}})}]{cordoba2002growth}%
  \BibitemOpen
  \bibfield  {author} {\bibinfo {author} {\bibfnamefont {D.}~\bibnamefont
  {C{\'o}rdoba}}\ and\ \bibinfo {author} {\bibfnamefont {C.}~\bibnamefont
  {Fefferman}},\ }\href@noop {} {\bibfield  {journal} {\bibinfo  {journal} {J.
  Am. Math. Soc.}\ }\textbf {\bibinfo {volume} {15}},\ \bibinfo {pages} {665}
  (\bibinfo {year} {2002}{\natexlab{a}})}\BibitemShut {NoStop}%
\bibitem [{\citenamefont {C{\'o}rdoba}\ and\ \citenamefont
  {Fefferman}(2002{\natexlab{b}})}]{cordoba2002scalars}%
  \BibitemOpen
  \bibfield  {author} {\bibinfo {author} {\bibfnamefont {D.}~\bibnamefont
  {C{\'o}rdoba}}\ and\ \bibinfo {author} {\bibfnamefont {C.}~\bibnamefont
  {Fefferman}},\ }\href@noop {} {\bibfield  {journal} {\bibinfo  {journal}
  {Comm. Pure Appl. Math.}\ }\textbf {\bibinfo {volume} {55}},\ \bibinfo
  {pages} {255} (\bibinfo {year} {2002}{\natexlab{b}})}\BibitemShut {NoStop}%
\bibitem [{\citenamefont {C{\'o}rdoba}\ and\ \citenamefont
  {C{\'o}rdoba}(2004)}]{cordoba2004maximum}%
  \BibitemOpen
  \bibfield  {author} {\bibinfo {author} {\bibfnamefont {A.}~\bibnamefont
  {C{\'o}rdoba}}\ and\ \bibinfo {author} {\bibfnamefont {D.}~\bibnamefont
  {C{\'o}rdoba}},\ }\href@noop {} {\bibfield  {journal} {\bibinfo  {journal}
  {Comm. Math. Phys.}\ }\textbf {\bibinfo {volume} {249}},\ \bibinfo {pages}
  {511} (\bibinfo {year} {2004})}\BibitemShut {NoStop}%
\bibitem [{\citenamefont {Rodrigo}\ and\ \citenamefont
  {Fefferman}(2004)}]{rodrigo2004vortex}%
  \BibitemOpen
  \bibfield  {author} {\bibinfo {author} {\bibfnamefont {J.}~\bibnamefont
  {Rodrigo}}\ and\ \bibinfo {author} {\bibfnamefont {C.}~\bibnamefont
  {Fefferman}},\ }\href@noop {} {\bibfield  {journal} {\bibinfo  {journal}
  {Proc. Natl. Acad. Sci.}\ ,\ \bibinfo {pages} {2684}} (\bibinfo {year}
  {2004})}\BibitemShut {NoStop}%
\bibitem [{\citenamefont {C{\'o}rdoba}\ \emph {et~al.}(2005)\citenamefont
  {C{\'o}rdoba}, \citenamefont {Fontelos}, \citenamefont {Mancho},\ and\
  \citenamefont {Rodrigo}}]{cordoba2005evidence}%
  \BibitemOpen
  \bibfield  {author} {\bibinfo {author} {\bibfnamefont {D.}~\bibnamefont
  {C{\'o}rdoba}}, \bibinfo {author} {\bibfnamefont {M.}~\bibnamefont
  {Fontelos}}, \bibinfo {author} {\bibfnamefont {A.}~\bibnamefont {Mancho}}, \
  and\ \bibinfo {author} {\bibfnamefont {J.}~\bibnamefont {Rodrigo}},\
  }\href@noop {} {\bibfield  {journal} {\bibinfo  {journal} {Proc. Natl. Acad.
  Sci.}\ }\textbf {\bibinfo {volume} {102}},\ \bibinfo {pages} {5949} (\bibinfo
  {year} {2005})}\BibitemShut {NoStop}%
\bibitem [{\citenamefont {Rodrigo}(2005)}]{rodrigo2005evolution}%
  \BibitemOpen
  \bibfield  {author} {\bibinfo {author} {\bibfnamefont {J.}~\bibnamefont
  {Rodrigo}},\ }\href@noop {} {\bibfield  {journal} {\bibinfo  {journal} {Comm.
  Pure Appl. Math.}\ }\textbf {\bibinfo {volume} {58}},\ \bibinfo {pages} {821}
  (\bibinfo {year} {2005})}\BibitemShut {NoStop}%
\bibitem [{\citenamefont {Wu}(2005)}]{wu2005solutions}%
  \BibitemOpen
  \bibfield  {author} {\bibinfo {author} {\bibfnamefont {J.}~\bibnamefont
  {Wu}},\ }\href@noop {} {\bibfield  {journal} {\bibinfo  {journal} {Nonlinear
  Anal.}\ }\textbf {\bibinfo {volume} {62}},\ \bibinfo {pages} {579} (\bibinfo
  {year} {2005})}\BibitemShut {NoStop}%
\bibitem [{\citenamefont {Deng}\ \emph {et~al.}(2006)\citenamefont {Deng},
  \citenamefont {Hou}, \citenamefont {Li},\ and\ \citenamefont
  {Yu}}]{deng2006level}%
  \BibitemOpen
  \bibfield  {author} {\bibinfo {author} {\bibfnamefont {J.}~\bibnamefont
  {Deng}}, \bibinfo {author} {\bibfnamefont {T.}~\bibnamefont {Hou}}, \bibinfo
  {author} {\bibfnamefont {R.}~\bibnamefont {Li}}, \ and\ \bibinfo {author}
  {\bibfnamefont {X.}~\bibnamefont {Yu}},\ }\href@noop {} {\bibfield  {journal}
  {\bibinfo  {journal} {Methods Appl. Anal.}\ }\textbf {\bibinfo {volume}
  {13}},\ \bibinfo {pages} {157} (\bibinfo {year} {2006})}\BibitemShut
  {NoStop}%
\bibitem [{\citenamefont {Dong}\ and\ \citenamefont
  {Li}(2008)}]{dong2008finite}%
  \BibitemOpen
  \bibfield  {author} {\bibinfo {author} {\bibfnamefont {H.}~\bibnamefont
  {Dong}}\ and\ \bibinfo {author} {\bibfnamefont {D.}~\bibnamefont {Li}},\
  }\href@noop {} {\bibfield  {journal} {\bibinfo  {journal} {Proc. Am. Math.
  Soc.}\ }\textbf {\bibinfo {volume} {136}},\ \bibinfo {pages} {2555} (\bibinfo
  {year} {2008})}\BibitemShut {NoStop}%
\bibitem [{\citenamefont {Ju}(2006)}]{ju2006geometric}%
  \BibitemOpen
  \bibfield  {author} {\bibinfo {author} {\bibfnamefont {N.}~\bibnamefont
  {Ju}},\ }\href@noop {} {\bibfield  {journal} {\bibinfo  {journal} {J. Diff.
  Eq.}\ }\textbf {\bibinfo {volume} {226}},\ \bibinfo {pages} {54} (\bibinfo
  {year} {2006})}\BibitemShut {NoStop}%
\bibitem [{\citenamefont {Li}(2009)}]{li2009existence}%
  \BibitemOpen
  \bibfield  {author} {\bibinfo {author} {\bibfnamefont {D.}~\bibnamefont
  {Li}},\ }\href@noop {} {\bibfield  {journal} {\bibinfo  {journal}
  {Nonlinearity}\ }\textbf {\bibinfo {volume} {22}},\ \bibinfo {pages} {1639}
  (\bibinfo {year} {2009})}\BibitemShut {NoStop}%
\bibitem [{\citenamefont
  {Marchand}(2008{\natexlab{a}})}]{marchand2008existence}%
  \BibitemOpen
  \bibfield  {author} {\bibinfo {author} {\bibfnamefont {F.}~\bibnamefont
  {Marchand}},\ }\href@noop {} {\bibfield  {journal} {\bibinfo  {journal}
  {Comm. Math. Phys.}\ }\textbf {\bibinfo {volume} {277}},\ \bibinfo {pages}
  {45} (\bibinfo {year} {2008}{\natexlab{a}})}\BibitemShut {NoStop}%
\bibitem [{\citenamefont {Marchand}(2008{\natexlab{b}})}]{marchand2008weak}%
  \BibitemOpen
  \bibfield  {author} {\bibinfo {author} {\bibfnamefont {F.}~\bibnamefont
  {Marchand}},\ }\href@noop {} {\bibfield  {journal} {\bibinfo  {journal}
  {Physica D}\ }\textbf {\bibinfo {volume} {237}},\ \bibinfo {pages} {1346}
  (\bibinfo {year} {2008}{\natexlab{b}})}\BibitemShut {NoStop}%
\bibitem [{\citenamefont {Scott}(2011)}]{scott2011scenario}%
  \BibitemOpen
  \bibfield  {author} {\bibinfo {author} {\bibfnamefont {R.}~\bibnamefont
  {Scott}},\ }\href@noop {} {\bibfield  {journal} {\bibinfo  {journal} {J.
  Fluid Mech.}\ }\textbf {\bibinfo {volume} {687}},\ \bibinfo {pages} {492}
  (\bibinfo {year} {2011})}\BibitemShut {NoStop}%
\bibitem [{\citenamefont {Constantin}\ \emph {et~al.}(2012)\citenamefont
  {Constantin}, \citenamefont {Lai}, \citenamefont {Sharma}, \citenamefont
  {Tseng},\ and\ \citenamefont {Wu}}]{constantin2012new}%
  \BibitemOpen
  \bibfield  {author} {\bibinfo {author} {\bibfnamefont {P.}~\bibnamefont
  {Constantin}}, \bibinfo {author} {\bibfnamefont {M.-C.}\ \bibnamefont {Lai}},
  \bibinfo {author} {\bibfnamefont {R.}~\bibnamefont {Sharma}}, \bibinfo
  {author} {\bibfnamefont {Y.-H.}\ \bibnamefont {Tseng}}, \ and\ \bibinfo
  {author} {\bibfnamefont {J.}~\bibnamefont {Wu}},\ }\href@noop {} {\bibfield
  {journal} {\bibinfo  {journal} {J. Sci. Comp.}\ }\textbf {\bibinfo {volume}
  {50}},\ \bibinfo {pages} {1} (\bibinfo {year} {2012})}\BibitemShut {NoStop}%
\bibitem [{\citenamefont {Scott}\ and\ \citenamefont
  {Dritschel}(2014)}]{scott2014numerical}%
  \BibitemOpen
  \bibfield  {author} {\bibinfo {author} {\bibfnamefont {R.}~\bibnamefont
  {Scott}}\ and\ \bibinfo {author} {\bibfnamefont {D.}~\bibnamefont
  {Dritschel}},\ }\href@noop {} {\bibfield  {journal} {\bibinfo  {journal}
  {Phys. Rev. Lett.}\ }\textbf {\bibinfo {volume} {112}},\ \bibinfo {pages}
  {144505} (\bibinfo {year} {2014})}\BibitemShut {NoStop}%
\bibitem [{\citenamefont {Ohkitani}(2012)}]{ohkitani2012asymptotics}%
  \BibitemOpen
  \bibfield  {author} {\bibinfo {author} {\bibfnamefont {K.}~\bibnamefont
  {Ohkitani}},\ }\href@noop {} {\bibfield  {journal} {\bibinfo  {journal}
  {Phys. Fluids}\ }\textbf {\bibinfo {volume} {24}},\ \bibinfo {pages} {095101}
  (\bibinfo {year} {2012})}\BibitemShut {NoStop}%
\bibitem [{\citenamefont {Hoyer}\ and\ \citenamefont
  {Sadourny}(1982)}]{hoyer1982closure}%
  \BibitemOpen
  \bibfield  {author} {\bibinfo {author} {\bibfnamefont {J.-M.}\ \bibnamefont
  {Hoyer}}\ and\ \bibinfo {author} {\bibfnamefont {R.}~\bibnamefont
  {Sadourny}},\ }\href@noop {} {\bibfield  {journal} {\bibinfo  {journal} {J.
  Atmos. Sci.}\ }\textbf {\bibinfo {volume} {39}},\ \bibinfo {pages} {707}
  (\bibinfo {year} {1982})}\BibitemShut {NoStop}%
\bibitem [{\citenamefont {Helmholtz}(1858)}]{helmholtz1858}%
  \BibitemOpen
  \bibfield  {author} {\bibinfo {author} {\bibfnamefont {H.}~\bibnamefont
  {Helmholtz}},\ }\href@noop {} {\bibfield  {journal} {\bibinfo  {journal} {J.
  Reine Angew. Math.}\ }\textbf {\bibinfo {volume} {55}},\ \bibinfo {pages}
  {25} (\bibinfo {year} {1858})}\BibitemShut {NoStop}%
\bibitem [{\citenamefont {Aref}(2007)}]{aref2007point}%
  \BibitemOpen
  \bibfield  {author} {\bibinfo {author} {\bibfnamefont {H.}~\bibnamefont
  {Aref}},\ }\href@noop {} {\bibfield  {journal} {\bibinfo  {journal} {J. Math.
  Phys.}\ }\textbf {\bibinfo {volume} {48}},\ \bibinfo {pages} {065401}
  (\bibinfo {year} {2007})}\BibitemShut {NoStop}%
\bibitem [{\citenamefont {Marchioro}\ and\ \citenamefont
  {Pulvirenti}(2012)}]{marchioro2012mathematical}%
  \BibitemOpen
  \bibfield  {author} {\bibinfo {author} {\bibfnamefont {C.}~\bibnamefont
  {Marchioro}}\ and\ \bibinfo {author} {\bibfnamefont {M.}~\bibnamefont
  {Pulvirenti}},\ }\href@noop {} {\emph {\bibinfo {title} {Mathematical theory
  of incompressible nonviscous fluids}}},\ Vol.~\bibinfo {volume} {96}\
  (\bibinfo  {publisher} {Springer Science \& Business Media},\ \bibinfo {year}
  {2012})\BibitemShut {NoStop}%
\bibitem [{\citenamefont {Newton}(2013)}]{newton2013n}%
  \BibitemOpen
  \bibfield  {author} {\bibinfo {author} {\bibfnamefont {P.}~\bibnamefont
  {Newton}},\ }\href@noop {} {\emph {\bibinfo {title} {The {N}-vortex problem:
  analytical techniques}}},\ Vol.\ \bibinfo {volume} {145}\ (\bibinfo
  {publisher} {Springer Science \& Business Media},\ \bibinfo {year}
  {2013})\BibitemShut {NoStop}%
\bibitem [{\citenamefont {Synge}(1949)}]{synge1949motion}%
  \BibitemOpen
  \bibfield  {author} {\bibinfo {author} {\bibfnamefont {J.}~\bibnamefont
  {Synge}},\ }\href@noop {} {\bibfield  {journal} {\bibinfo  {journal} {Can. J.
  Math.}\ }\textbf {\bibinfo {volume} {1}},\ \bibinfo {pages} {257} (\bibinfo
  {year} {1949})}\BibitemShut {NoStop}%
\bibitem [{\citenamefont {Aref}(1979)}]{aref1979motion}%
  \BibitemOpen
  \bibfield  {author} {\bibinfo {author} {\bibfnamefont {H.}~\bibnamefont
  {Aref}},\ }\href@noop {} {\bibfield  {journal} {\bibinfo  {journal} {Phys.
  Fluids}\ }\textbf {\bibinfo {volume} {22}},\ \bibinfo {pages} {393} (\bibinfo
  {year} {1979})}\BibitemShut {NoStop}%
\bibitem [{\citenamefont {Novikov}\ and\ \citenamefont
  {Sedov}(1979)}]{novikov1979vortex}%
  \BibitemOpen
  \bibfield  {author} {\bibinfo {author} {\bibfnamefont {E.}~\bibnamefont
  {Novikov}}\ and\ \bibinfo {author} {\bibfnamefont {Y.}~\bibnamefont
  {Sedov}},\ }\href@noop {} {\bibfield  {journal} {\bibinfo  {journal} {Zh.
  Eksp. Teor. Fiz.}\ }\textbf {\bibinfo {volume} {77}},\ \bibinfo {pages} {588}
  (\bibinfo {year} {1979})}\BibitemShut {NoStop}%
\bibitem [{\citenamefont {O'Neil}(1987)}]{o1987stationary}%
  \BibitemOpen
  \bibfield  {author} {\bibinfo {author} {\bibfnamefont {K.}~\bibnamefont
  {O'Neil}},\ }\href@noop {} {\bibfield  {journal} {\bibinfo  {journal} {Trans.
  Am. Math. Soc.}\ }\textbf {\bibinfo {volume} {302}},\ \bibinfo {pages} {383}
  (\bibinfo {year} {1987})}\BibitemShut {NoStop}%
\bibitem [{\citenamefont {Aref}(2010)}]{aref2010self}%
  \BibitemOpen
  \bibfield  {author} {\bibinfo {author} {\bibfnamefont {H.}~\bibnamefont
  {Aref}},\ }\href@noop {} {\bibfield  {journal} {\bibinfo  {journal} {Phys.
  Fluids}\ }\textbf {\bibinfo {volume} {22}},\ \bibinfo {pages} {057104}
  (\bibinfo {year} {2010})}\BibitemShut {NoStop}%
\bibitem [{\citenamefont {Sakajo}(2008)}]{sakajo2008non}%
  \BibitemOpen
  \bibfield  {author} {\bibinfo {author} {\bibfnamefont {T.}~\bibnamefont
  {Sakajo}},\ }\href@noop {} {\bibfield  {journal} {\bibinfo  {journal} {Phys.
  Rev. E}\ }\textbf {\bibinfo {volume} {78}},\ \bibinfo {pages} {016312}
  (\bibinfo {year} {2008})}\BibitemShut {NoStop}%
\bibitem [{\citenamefont {Tavantzis}\ and\ \citenamefont
  {Ting}(1988)}]{tavantzis1988dynamics}%
  \BibitemOpen
  \bibfield  {author} {\bibinfo {author} {\bibfnamefont {J.}~\bibnamefont
  {Tavantzis}}\ and\ \bibinfo {author} {\bibfnamefont {L.}~\bibnamefont
  {Ting}},\ }\href@noop {} {\bibfield  {journal} {\bibinfo  {journal} {Phys.
  Fluids}\ }\textbf {\bibinfo {volume} {31}},\ \bibinfo {pages} {1392}
  (\bibinfo {year} {1988})}\BibitemShut {NoStop}%
\bibitem [{\citenamefont {Kimura}\ \emph {et~al.}(1990)\citenamefont {Kimura},
  \citenamefont {Zawadzki},\ and\ \citenamefont {Aref}}]{kimura1990vortex}%
  \BibitemOpen
  \bibfield  {author} {\bibinfo {author} {\bibfnamefont {Y.}~\bibnamefont
  {Kimura}}, \bibinfo {author} {\bibfnamefont {I.}~\bibnamefont {Zawadzki}}, \
  and\ \bibinfo {author} {\bibfnamefont {H.}~\bibnamefont {Aref}},\ }\href@noop
  {} {\bibfield  {journal} {\bibinfo  {journal} {Phys. Fluids A}\ }\textbf
  {\bibinfo {volume} {2}},\ \bibinfo {pages} {214} (\bibinfo {year}
  {1990})}\BibitemShut {NoStop}%
\bibitem [{\citenamefont {Kimura}(1990)}]{kimura1990parametric}%
  \BibitemOpen
  \bibfield  {author} {\bibinfo {author} {\bibfnamefont {Y.}~\bibnamefont
  {Kimura}},\ }\href@noop {} {\bibfield  {journal} {\bibinfo  {journal}
  {Physica D}\ }\textbf {\bibinfo {volume} {46}},\ \bibinfo {pages} {439}
  (\bibinfo {year} {1990})}\BibitemShut {NoStop}%
\bibitem [{\citenamefont {Kimura}(1991)}]{kimura1991vortex}%
  \BibitemOpen
  \bibfield  {author} {\bibinfo {author} {\bibfnamefont {Y.}~\bibnamefont
  {Kimura}},\ }\href@noop {} {\bibfield  {journal} {\bibinfo  {journal}
  {Physica D}\ }\textbf {\bibinfo {volume} {51}},\ \bibinfo {pages} {512}
  (\bibinfo {year} {1991})}\BibitemShut {NoStop}%
\bibitem [{\citenamefont {Leoncini}\ \emph {et~al.}(2000)\citenamefont
  {Leoncini}, \citenamefont {Kuznetsov},\ and\ \citenamefont
  {Zaslavsky}}]{leoncini2000motion}%
  \BibitemOpen
  \bibfield  {author} {\bibinfo {author} {\bibfnamefont {X.}~\bibnamefont
  {Leoncini}}, \bibinfo {author} {\bibfnamefont {L.}~\bibnamefont {Kuznetsov}},
  \ and\ \bibinfo {author} {\bibfnamefont {G.}~\bibnamefont {Zaslavsky}},\
  }\href@noop {} {\bibfield  {journal} {\bibinfo  {journal} {Phys. Fluids}\
  }\textbf {\bibinfo {volume} {12}},\ \bibinfo {pages} {1911} (\bibinfo {year}
  {2000})}\BibitemShut {NoStop}%
\bibitem [{\citenamefont {Leoncini}\ \emph {et~al.}(2001)\citenamefont
  {Leoncini}, \citenamefont {Kuznetsov},\ and\ \citenamefont
  {Zaslavsky}}]{leoncini2001chaotic}%
  \BibitemOpen
  \bibfield  {author} {\bibinfo {author} {\bibfnamefont {X.}~\bibnamefont
  {Leoncini}}, \bibinfo {author} {\bibfnamefont {L.}~\bibnamefont {Kuznetsov}},
  \ and\ \bibinfo {author} {\bibfnamefont {G.}~\bibnamefont {Zaslavsky}},\
  }\href@noop {} {\bibfield  {journal} {\bibinfo  {journal} {Phys. Rev. E}\
  }\textbf {\bibinfo {volume} {63}},\ \bibinfo {pages} {036224} (\bibinfo
  {year} {2001})}\BibitemShut {NoStop}%
\bibitem [{\citenamefont {Hern{\'a}ndez-Garduno}\ and\ \citenamefont
  {Lacomba}(2007)}]{hernandez2007collisions}%
  \BibitemOpen
  \bibfield  {author} {\bibinfo {author} {\bibfnamefont {A.}~\bibnamefont
  {Hern{\'a}ndez-Garduno}}\ and\ \bibinfo {author} {\bibfnamefont
  {E.}~\bibnamefont {Lacomba}},\ }\href@noop {} {\bibfield  {journal} {\bibinfo
   {journal} {J. Math. Fluid Mech.}\ }\textbf {\bibinfo {volume} {9}},\
  \bibinfo {pages} {75} (\bibinfo {year} {2007})}\BibitemShut {NoStop}%
\bibitem [{\citenamefont {Sakajo}(2012)}]{sakajo2012instantaneous}%
  \BibitemOpen
  \bibfield  {author} {\bibinfo {author} {\bibfnamefont {T.}~\bibnamefont
  {Sakajo}},\ }\href@noop {} {\bibfield  {journal} {\bibinfo  {journal} {J.
  Fluid Mech.}\ }\textbf {\bibinfo {volume} {702}},\ \bibinfo {pages} {188}
  (\bibinfo {year} {2012})}\BibitemShut {NoStop}%
\bibitem [{\citenamefont {Sakajo}(2013)}]{sakajo2013anomalous}%
  \BibitemOpen
  \bibfield  {author} {\bibinfo {author} {\bibfnamefont {T.}~\bibnamefont
  {Sakajo}},\ }in\ \href@noop {} {\emph {\bibinfo {booktitle} {Emerging Topics
  on Differential Equations and Their Applications}}}\ (\bibinfo  {publisher}
  {World Scientific},\ \bibinfo {year} {2013})\ pp.\ \bibinfo {pages}
  {155--169}\BibitemShut {NoStop}%
\bibitem [{\citenamefont {Kudela}(2014)}]{kudela2014collapse}%
  \BibitemOpen
  \bibfield  {author} {\bibinfo {author} {\bibfnamefont {H.}~\bibnamefont
  {Kudela}},\ }\href@noop {} {\bibfield  {journal} {\bibinfo  {journal} {Fluid
  Dyn. Res.}\ }\textbf {\bibinfo {volume} {46}},\ \bibinfo {pages} {031414}
  (\bibinfo {year} {2014})}\BibitemShut {NoStop}%
\bibitem [{\citenamefont {Kirchhoff}(1876)}]{kirchhoff1876vorlesungen}%
  \BibitemOpen
  \bibfield  {author} {\bibinfo {author} {\bibfnamefont {G.}~\bibnamefont
  {Kirchhoff}},\ }\href@noop {} {\emph {\bibinfo {title} {{Vorlesungen {\"u}ber
  mathematische Physik: Mechanik}}}},\ Vol.~\bibinfo {volume} {1}\ (\bibinfo
  {publisher} {Teubner},\ \bibinfo {year} {1876})\BibitemShut {NoStop}%
\bibitem [{\citenamefont {Nambu}(1973)}]{Nambu1973}%
  \BibitemOpen
  \bibfield  {author} {\bibinfo {author} {\bibfnamefont {Y.}~\bibnamefont
  {Nambu}},\ }\href@noop {} {\bibfield  {journal} {\bibinfo  {journal} {Phys.
  Rev. D}\ }\textbf {\bibinfo {volume} {7}},\ \bibinfo {pages} {2405} (\bibinfo
  {year} {1973})}\BibitemShut {NoStop}%
\bibitem [{\citenamefont {Takhtajan}(1994)}]{Takhtajan1994}%
  \BibitemOpen
  \bibfield  {author} {\bibinfo {author} {\bibfnamefont {L.}~\bibnamefont
  {Takhtajan}},\ }\href@noop {} {\bibfield  {journal} {\bibinfo  {journal}
  {Commun. Math. Phys.}\ }\textbf {\bibinfo {volume} {160}},\ \bibinfo {pages}
  {295} (\bibinfo {year} {1994})}\BibitemShut {NoStop}%
\bibitem [{\citenamefont {N{\'e}vir}\ and\ \citenamefont
  {Blender}(1994)}]{NevirBlender1994}%
  \BibitemOpen
  \bibfield  {author} {\bibinfo {author} {\bibfnamefont {P.}~\bibnamefont
  {N{\'e}vir}}\ and\ \bibinfo {author} {\bibfnamefont {R.}~\bibnamefont
  {Blender}},\ }\href@noop {} {\bibfield  {journal} {\bibinfo  {journal}
  {Contrib. to Atmosph. Phys.}\ }\textbf {\bibinfo {volume} {67}},\ \bibinfo
  {pages} {133} (\bibinfo {year} {1994})}\BibitemShut {NoStop}%
\bibitem [{\citenamefont {Blender}\ and\ \citenamefont
  {Badin}(2015)}]{blender2015hydrodynamic}%
  \BibitemOpen
  \bibfield  {author} {\bibinfo {author} {\bibfnamefont {R.}~\bibnamefont
  {Blender}}\ and\ \bibinfo {author} {\bibfnamefont {G.}~\bibnamefont
  {Badin}},\ }\href@noop {} {\bibfield  {journal} {\bibinfo  {journal} {J.
  Phys. A: Math. Theor.}\ }\textbf {\bibinfo {volume} {48}},\ \bibinfo {pages}
  {105501} (\bibinfo {year} {2015})}\BibitemShut {NoStop}%
\bibitem [{\citenamefont {N\'{e}vir}\ and\ \citenamefont
  {Blender}(1993)}]{NevirBlender1993}%
  \BibitemOpen
  \bibfield  {author} {\bibinfo {author} {\bibfnamefont {P.}~\bibnamefont
  {N\'{e}vir}}\ and\ \bibinfo {author} {\bibfnamefont {R.}~\bibnamefont
  {Blender}},\ }\href@noop {} {\bibfield  {journal} {\bibinfo  {journal} {J.
  Phys. A: Math. Gen.}\ }\textbf {\bibinfo {volume} {26}},\ \bibinfo {pages}
  {1189} (\bibinfo {year} {1993})}\BibitemShut {NoStop}%
\bibitem [{\citenamefont {Salmon}(2005)}]{Salmon2005}%
  \BibitemOpen
  \bibfield  {author} {\bibinfo {author} {\bibfnamefont {R.}~\bibnamefont
  {Salmon}},\ }\href@noop {} {\bibfield  {journal} {\bibinfo  {journal}
  {Nonlinearity}\ }\textbf {\bibinfo {volume} {18}},\ \bibinfo {pages} {1}
  (\bibinfo {year} {2005})}\BibitemShut {NoStop}%
\bibitem [{\citenamefont {Bridges}\ and\ \citenamefont
  {Reich}(2006)}]{bridges2006numerical}%
  \BibitemOpen
  \bibfield  {author} {\bibinfo {author} {\bibfnamefont {T.~J.}\ \bibnamefont
  {Bridges}}\ and\ \bibinfo {author} {\bibfnamefont {S.}~\bibnamefont
  {Reich}},\ }\href@noop {} {\bibfield  {journal} {\bibinfo  {journal} {J.
  Phys. A: Math. Theor.}\ }\textbf {\bibinfo {volume} {39}},\ \bibinfo {pages}
  {5287} (\bibinfo {year} {2006})}\BibitemShut {NoStop}%
\bibitem [{\citenamefont {Salmon}(2007)}]{Salmon2007}%
  \BibitemOpen
  \bibfield  {author} {\bibinfo {author} {\bibfnamefont {R.}~\bibnamefont
  {Salmon}},\ }\href@noop {} {\bibfield  {journal} {\bibinfo  {journal} {J.
  Atmos. Sci.}\ }\textbf {\bibinfo {volume} {64}},\ \bibinfo {pages} {515}
  (\bibinfo {year} {2007})}\BibitemShut {NoStop}%
\bibitem [{\citenamefont {Bihlo}(2008)}]{Bihlo2008}%
  \BibitemOpen
  \bibfield  {author} {\bibinfo {author} {\bibfnamefont {A.}~\bibnamefont
  {Bihlo}},\ }\href@noop {} {\bibfield  {journal} {\bibinfo  {journal} {J.
  Phys. A: Math. Theor.}\ }\textbf {\bibinfo {volume} {41}},\ \bibinfo {pages}
  {292001} (\bibinfo {year} {2008})}\BibitemShut {NoStop}%
\bibitem [{\citenamefont {Gassmann}\ and\ \citenamefont
  {Herzog}(2008)}]{GassmannHerzog2008}%
  \BibitemOpen
  \bibfield  {author} {\bibinfo {author} {\bibfnamefont {A.}~\bibnamefont
  {Gassmann}}\ and\ \bibinfo {author} {\bibfnamefont {H.-J.}\ \bibnamefont
  {Herzog}},\ }\href@noop {} {\bibfield  {journal} {\bibinfo  {journal} {Quart.
  J. Roy. Meteorol. Soc.}\ }\textbf {\bibinfo {volume} {134}},\ \bibinfo
  {pages} {1597} (\bibinfo {year} {2008})}\BibitemShut {NoStop}%
\bibitem [{\citenamefont {N\'{e}vir}\ and\ \citenamefont
  {Sommer}(2009)}]{NevirSommer2009}%
  \BibitemOpen
  \bibfield  {author} {\bibinfo {author} {\bibfnamefont {P.}~\bibnamefont
  {N\'{e}vir}}\ and\ \bibinfo {author} {\bibfnamefont {M.}~\bibnamefont
  {Sommer}},\ }\href@noop {} {\bibfield  {journal} {\bibinfo  {journal} {J.
  Atmos. Sci.}\ }\textbf {\bibinfo {volume} {66}},\ \bibinfo {pages} {2073}
  (\bibinfo {year} {2009})}\BibitemShut {NoStop}%
\bibitem [{\citenamefont {Sommer}\ and\ \citenamefont
  {N\'{e}vir}(2009)}]{SommerNevir2009}%
  \BibitemOpen
  \bibfield  {author} {\bibinfo {author} {\bibfnamefont {M.}~\bibnamefont
  {Sommer}}\ and\ \bibinfo {author} {\bibfnamefont {P.}~\bibnamefont
  {N\'{e}vir}},\ }\href@noop {} {\bibfield  {journal} {\bibinfo  {journal}
  {Quart. J. Roy. Meteorol. Soc.}\ }\textbf {\bibinfo {volume} {135}},\
  \bibinfo {pages} {485} (\bibinfo {year} {2009})}\BibitemShut {NoStop}%
\bibitem [{\citenamefont {Salazar}\ and\ \citenamefont
  {Kurgansky}(2010)}]{SalazarKurgansky2010}%
  \BibitemOpen
  \bibfield  {author} {\bibinfo {author} {\bibfnamefont {R.}~\bibnamefont
  {Salazar}}\ and\ \bibinfo {author} {\bibfnamefont {M.~V.}\ \bibnamefont
  {Kurgansky}},\ }\href@noop {} {\bibfield  {journal} {\bibinfo  {journal} {J.
  Phys. A}\ }\textbf {\bibinfo {volume} {43}},\ \bibinfo {pages} {305501}
  (\bibinfo {year} {2010})}\BibitemShut {NoStop}%
\bibitem [{\citenamefont {Blender}\ and\ \citenamefont
  {Lucarini}(2013)}]{BlenderLucarini2013}%
  \BibitemOpen
  \bibfield  {author} {\bibinfo {author} {\bibfnamefont {R.}~\bibnamefont
  {Blender}}\ and\ \bibinfo {author} {\bibfnamefont {V.}~\bibnamefont
  {Lucarini}},\ }\href@noop {} {\bibfield  {journal} {\bibinfo  {journal}
  {Physica D}\ }\textbf {\bibinfo {volume} {243}},\ \bibinfo {pages} {86}
  (\bibinfo {year} {2013})}\BibitemShut {NoStop}%
\bibitem [{\citenamefont {Blender}\ and\ \citenamefont
  {Badin}(2017{\natexlab{a}})}]{blender2017viscous}%
  \BibitemOpen
  \bibfield  {author} {\bibinfo {author} {\bibfnamefont {R.}~\bibnamefont
  {Blender}}\ and\ \bibinfo {author} {\bibfnamefont {G.}~\bibnamefont
  {Badin}},\ }\href@noop {} {\bibfield  {journal} {\bibinfo  {journal} {Eur.
  Phys. J. Plus}\ }\textbf {\bibinfo {volume} {132}},\ \bibinfo {pages} {1}
  (\bibinfo {year} {2017}{\natexlab{a}})}\BibitemShut {NoStop}%
\bibitem [{\citenamefont {Blender}\ and\ \citenamefont
  {Badin}(2017{\natexlab{b}})}]{blender2017construction}%
  \BibitemOpen
  \bibfield  {author} {\bibinfo {author} {\bibfnamefont {R.}~\bibnamefont
  {Blender}}\ and\ \bibinfo {author} {\bibfnamefont {G.}~\bibnamefont
  {Badin}},\ }\href@noop {} {\bibfield  {journal} {\bibinfo  {journal}
  {Fluids}\ }\textbf {\bibinfo {volume} {2}},\ \bibinfo {pages} {24} (\bibinfo
  {year} {2017}{\natexlab{b}})}\BibitemShut {NoStop}%
\bibitem [{\citenamefont {Makhaldiani}(2007)}]{Makhaldiani07}%
  \BibitemOpen
  \bibfield  {author} {\bibinfo {author} {\bibfnamefont {N.}~\bibnamefont
  {Makhaldiani}},\ }\href@noop {} {\bibfield  {journal} {\bibinfo  {journal}
  {Phys. Atom. Nucl.}\ }\textbf {\bibinfo {volume} {70}},\ \bibinfo {pages}
  {567} (\bibinfo {year} {2007})}\BibitemShut {NoStop}%
\bibitem [{\citenamefont {Makhaldiani}(2012)}]{Makhaldiani12}%
  \BibitemOpen
  \bibfield  {author} {\bibinfo {author} {\bibfnamefont {N.}~\bibnamefont
  {Makhaldiani}},\ }\href@noop {} {\bibfield  {journal} {\bibinfo  {journal}
  {Phys. Part. Nuclei}\ }\textbf {\bibinfo {volume} {43}},\ \bibinfo {pages}
  {703} (\bibinfo {year} {2012})}\BibitemShut {NoStop}%
\bibitem [{\citenamefont {M\"{u}ller}\ and\ \citenamefont
  {N\'{e}vir}(2014)}]{MuellerNevir14}%
  \BibitemOpen
  \bibfield  {author} {\bibinfo {author} {\bibfnamefont {A.}~\bibnamefont
  {M\"{u}ller}}\ and\ \bibinfo {author} {\bibfnamefont {P.}~\bibnamefont
  {N\'{e}vir}},\ }\href@noop {} {\bibfield  {journal} {\bibinfo  {journal} {J.
  Phys. A: Math. Theor.}\ }\textbf {\bibinfo {volume} {47}},\ \bibinfo {pages}
  {105201} (\bibinfo {year} {2014})}\BibitemShut {NoStop}%
\bibitem [{\citenamefont {Taylor}\ and\ \citenamefont
  {Llewellyn~Smith}(2016)}]{taylor2016dynamics}%
  \BibitemOpen
  \bibfield  {author} {\bibinfo {author} {\bibfnamefont {C.}~\bibnamefont
  {Taylor}}\ and\ \bibinfo {author} {\bibfnamefont {S.}~\bibnamefont
  {Llewellyn~Smith}},\ }\href@noop {} {\bibfield  {journal} {\bibinfo
  {journal} {Chaos}\ }\textbf {\bibinfo {volume} {26}},\ \bibinfo {pages}
  {113117} (\bibinfo {year} {2016})}\BibitemShut {NoStop}%
\bibitem [{\citenamefont {Lim}\ and\ \citenamefont
  {Majda}(2001)}]{lim2001point}%
  \BibitemOpen
  \bibfield  {author} {\bibinfo {author} {\bibfnamefont {C.}~\bibnamefont
  {Lim}}\ and\ \bibinfo {author} {\bibfnamefont {A.}~\bibnamefont {Majda}},\
  }\href@noop {} {\bibfield  {journal} {\bibinfo  {journal} {Geophys. Astro.
  Fluid Dyn.}\ }\textbf {\bibinfo {volume} {94}},\ \bibinfo {pages} {177}
  (\bibinfo {year} {2001})}\BibitemShut {NoStop}%
\bibitem [{\citenamefont {Iwayama}\ and\ \citenamefont
  {Watanabe}(2010)}]{Iwayama2010}%
  \BibitemOpen
  \bibfield  {author} {\bibinfo {author} {\bibfnamefont {T.}~\bibnamefont
  {Iwayama}}\ and\ \bibinfo {author} {\bibfnamefont {T.}~\bibnamefont
  {Watanabe}},\ }\href@noop {} {\bibfield  {journal} {\bibinfo  {journal}
  {Phys. Rev. E}\ }\textbf {\bibinfo {volume} {82}},\ \bibinfo {pages} {036307}
  (\bibinfo {year} {2010})}\BibitemShut {NoStop}%
\bibitem [{\citenamefont {Chapman}(1978)}]{chapman1978ideal}%
  \BibitemOpen
  \bibfield  {author} {\bibinfo {author} {\bibfnamefont {D.}~\bibnamefont
  {Chapman}},\ }\href@noop {} {\bibfield  {journal} {\bibinfo  {journal} {J.
  Math. Phys.}\ }\textbf {\bibinfo {volume} {19}},\ \bibinfo {pages} {1988}
  (\bibinfo {year} {1978})}\BibitemShut {NoStop}%
\bibitem [{\citenamefont {Gr{\"o}bli}(1877)}]{groblispezielle}%
  \BibitemOpen
  \bibfield  {author} {\bibinfo {author} {\bibfnamefont {W.}~\bibnamefont
  {Gr{\"o}bli}},\ }\href@noop {} {\bibfield  {journal} {\bibinfo  {journal}
  {Vierteljahrsschr. Natforsch. Ges. Zur.}\ }\textbf {\bibinfo {volume} {22}},\
  \bibinfo {pages} {37} (\bibinfo {year} {1877})}\BibitemShut {NoStop}%
\bibitem [{\citenamefont {Novikov}(1975)}]{novikov1975dynamics}%
  \BibitemOpen
  \bibfield  {author} {\bibinfo {author} {\bibfnamefont {E.}~\bibnamefont
  {Novikov}},\ }\href@noop {} {\bibfield  {journal} {\bibinfo  {journal} {Zh.
  Eksp. Teor. Fiz}\ }\textbf {\bibinfo {volume} {68}},\ \bibinfo {pages} {2}
  (\bibinfo {year} {1975})}\BibitemShut {NoStop}%
\bibitem [{\citenamefont {Vosbeek}\ \emph {et~al.}(1997)\citenamefont
  {Vosbeek}, \citenamefont {Van~Geffen}, \citenamefont {Meleshko},\ and\
  \citenamefont {Van~Heijst}}]{vosbeek1997collapse}%
  \BibitemOpen
  \bibfield  {author} {\bibinfo {author} {\bibfnamefont {P.}~\bibnamefont
  {Vosbeek}}, \bibinfo {author} {\bibfnamefont {J.}~\bibnamefont {Van~Geffen}},
  \bibinfo {author} {\bibfnamefont {V.}~\bibnamefont {Meleshko}}, \ and\
  \bibinfo {author} {\bibfnamefont {G.}~\bibnamefont {Van~Heijst}},\
  }\href@noop {} {\bibfield  {journal} {\bibinfo  {journal} {Phys. Fluids}\
  }\textbf {\bibinfo {volume} {9}},\ \bibinfo {pages} {3315} (\bibinfo {year}
  {1997})}\BibitemShut {NoStop}%
\bibitem [{\citenamefont {Badin}(2013)}]{badin2013surface}%
  \BibitemOpen
  \bibfield  {author} {\bibinfo {author} {\bibfnamefont {G.}~\bibnamefont
  {Badin}},\ }\href@noop {} {\bibfield  {journal} {\bibinfo  {journal}
  {Geophys. Astrophys. Fluid Dyn.}\ }\textbf {\bibinfo {volume} {107}},\
  \bibinfo {pages} {526} (\bibinfo {year} {2013})}\BibitemShut {NoStop}%
\bibitem [{\citenamefont {Ragone}\ and\ \citenamefont
  {Badin}(2016)}]{ragone2016study}%
  \BibitemOpen
  \bibfield  {author} {\bibinfo {author} {\bibfnamefont {F.}~\bibnamefont
  {Ragone}}\ and\ \bibinfo {author} {\bibfnamefont {G.}~\bibnamefont {Badin}},\
  }\href@noop {} {\bibfield  {journal} {\bibinfo  {journal} {J. Fluid Mech.}\
  }\textbf {\bibinfo {volume} {792}},\ \bibinfo {pages} {740} (\bibinfo {year}
  {2016})}\BibitemShut {NoStop}%
\end{thebibliography}

\end{document}